\documentclass{JFM-FLM_Au}

\usepackage{lineno}
\usepackage{amsmath,amssymb}
\usepackage{pdflscape}
\usepackage{array}
\usepackage{multirow}
\usepackage{subcaption}
\usepackage{float}
\usepackage{etoolbox}
\usepackage{hyperref}

\lefttitle{Y. Bimali, R. McCabe, C. Treacy, K. Khanal, E. Lo,  and M. N. Haji}
\righttitle{Journal of Fluid Mechanics}

\title{Matrix structure and convergence behaviour of the matched eigenfunction method for computing heave wave forces on generalized concentric bodies}

\author{Yinghui Bimali\aff{1,\textdagger},
Rebecca McCabe\aff{2,\textdagger}, 
Collin Treacy\aff{3},
Kapil Khanal\aff{4},
En Lo\aff{5,6},
\and 
Maha Haji\aff{2,3}}

\affiliation{
\aff{1} Dept. of Applied and Engineering Physics, Cornell Univ., 142 Sciences Dr., Ithaca, NY 14853, USA
\aff{2}Sibley Sch. of Mechanical \& Aerospace Engineering, Cornell Univ., 124 Hoy Rd, Ithaca, NY 14853, USA
\aff{3} Dept. of Mechanical Engineering, Univ. of Michigan, 2350 Hayward St., Ann Arbor, MI 48109, USA
\aff{4}Dept. of Systems Engineering, Cornell Univ., 136 Hoy Road, Ithaca, NY 14853, USA
\aff{5} Dept. of Environmental Engineering, Cornell Univ., 616 Thurston Ave, Ithaca, NY 14853, USA
\aff{6} Dept. of Engineering Science, Parks Road, Univ. of Oxford, OX1 3PJ, UK
\aff{\textdagger}These authors contributed equally to this work.
}

\corresau{Rebecca McCabe, \email{rgm222@cornell.edu}}

\BeforeBeginEnvironment{equation}{\par\vspace{-\baselineskip}\begin{linenomath*}}
\AfterEndEnvironment{equation}{\end{linenomath*}}
\BeforeBeginEnvironment{multline}{\par\vspace{-1.5\baselineskip}\begin{linenomath*}}
\AfterEndEnvironment{multline}{\end{linenomath*}}

\begin{document}
\maketitle
\begin{abstract}Structural survival of offshore structures is crucial for the growing marine economy.
Calculating the added mass, radiation damping, and excitation coefficients to quantify wave loads with the traditional boundary element method (BEM) presents a computational bottleneck. 
The matched eigenfunction expansion method (MEEM), a long-known but rarely-used alternative, offers computational benefits due to its semi-analytical nature. 
However, previous work fails to directly compare its accuracy and computational performance with BEM, leaving the extent of its utility unknown. 
Furthermore, the geometry-dependent convergence for cylindrical and slanted geometries has not yet been documented, making the method’s practicality for general geometries unclear. 
This paper presents a unifying MEEM framework for modelling an arbitrary number of fixed or heaving surface-piercing annular cylinders with continuous and radially-monotonic body profiles,
and explores the method’s block matrix structure, convergence behaviour, ability to accurately approximate slanted geometries, and computational advantages over the BEM solver Capytaine. 
The numerical experiments show that 
MEEM can compute hydrodynamic coefficients of slanted geometries within 5\% of Capytaine, even for angles as steep as $15^\circ$ from vertical.
Finally, 
MEEM can achieve 2\% convergence of its hydrodynamic coefficients an order of magnitude faster than Capytaine with a matrix size two orders of magnitude smaller, making it a computationally effective alternative to traditional BEM solvers.
These contributions enable hydrodynamic analysis of a broad range of shapes with increased speed and confidence, paving the way for future optimization studies to yield improved designs.

\end{abstract}

\begin{keywords}
Authors should not enter keywords on the manuscript, as these must be chosen by the author during the online submission process and will then be added during the typesetting process (see \href{https://www.cambridge.org/core/journals/journal-of-fluid-mechanics/information/list-of-keywords}{Keyword PDF} for the full list).  Other classifications will be added at the same time.
\end{keywords}


\section{Introduction}
\label{sec:Introduction}
Linear potential flow theory is a simplification of the Navier-Stokes equations widely used to model wave-structure interactions.
Early analytical solutions for the wave force on vertical cylindrical bodies under linear potential flow include solutions for infinite cylinders in infinite depth \citep{havelock_forces_1936} and bottom-mounted cylinders in finite depth \citep{fuchs_wave_1954}.
Both take advantage of the separability of the Laplace equation in cylindrical coordinates.
Fully analytical solutions are not available for truncated cylinders (those whose bottoms do not touch the sea floor) since it is impossible to analytically enforce continuity of potential between the water under the body and the water surrounding it.

\citet{yeung_added_1981} presents a semi-analytical method called the matched eigenfunction expansion method (MEEM) to address this challenge.
The solution in each fluid region is obtained analytically in terms of unknown coefficients, and these coefficients are found numerically by enforcing continuity of potential and radial velocity at region boundaries.
The method received thorough coverage in a recent analytical hydrodynamics textbook \citep{chatzigeorgiou2018analytical} and has been extended to cover increasingly complex concentric geometries, 
including two truncated cylinders \citep{mavrakos_hydrodynamic_2004,chau2012inertia}, three truncated cylinders \citep{zhang_performance_2024}, damping plates \citep{olaya_hydrodynamic_2015}, and many cylinders approximating complex concentric bodies \citep{kokkinowrachos_behaviour_1986}.

Despite this interest from hydrodynamics researchers, MEEM has not been widely adopted by the offshore engineering community, likely due to the complexity of its implementation 
and the rise of boundary element method (BEM) software packages that are easier to use and applicable to arbitrary geometries, albeit more computationally costly.
With the growing need for efficient hydrodynamic analysis tools to enable advanced design procedures like optimization, interest in MEEM has resurged, particularly in the offshore renewable energy space. 
\citet{pavlidou_novel_2022} performed multi-objective optimization for a two-body wave energy converter (WEC) using MEEM to compute the hydrodynamic coefficients of a cylinder. 
\citet{zhang_hydrodynamic_2016} optimized the shape of the wetted surface of a point absorber WEC using MEEM to model cylindrical, hemispherical, paraboloid, and conical geometries. 
\citet{mavrakos_hydrodynamic_2012} used MEEM to compute the first-order and mean second-order loads on a vertical axisymmetric oscillating water column WEC. 
However, the benefits of MEEM over traditional BEM solvers remain unclear, as the tradeoffs between accuracy and computational performance of the methods are seldom compared in the literature. 
Additionally, previous work does not directly address the convergence behaviour and angle-dependent discretization of slanted geometries, making its ability to accurately model more general geometries uncertain.

This work presents a unifying MEEM framework that arranges equations into an organized sparse matrix structure. 
This framework is used to model an arbitrary number of fixed or heaving surface-piercing annular cylinders with continuous radially-monotonic profiles. 
The method’s convergence behaviour, ability to accurately approximate slanted bodies, and computational advantages over the BEM solver Capytaine are explored. 
The authors build on their preliminary work \citep{mccabe_open-source_2024} and have developed an open-source Python package called \href{https://github.com/symbiotic-engineering/OpenFLASH}{\texttt{OpenFLASH}} (Flexible Library for Analytical \& Semi-analytical Hydrodynamics, \citep{best_openflash_2026,best_openflash-code_2026}) that is used for numerical studies herein.

Section~\ref{sec:mathematical-formulation} formulates the mathematics, unifies the aforementioned literature in a consistent notation, 
explains the method for an audience unfamiliar with analytical hydrodynamics, clarifies the block matrix structure and asymptotic behaviour for small and large frequencies, discusses numerical subtleties needed to avoid overflow and finite precision effects, and validates the results of OpenFLASH against those of Capytaine.
Section~\ref{sec:convergence} details the novel results of a convergence study and provides guidance on the number of terms required to achieve a desired accuracy for a given geometric configuration.
Section~\ref{sec:slant} extends the method to bodies with slanted (conical) surfaces, evaluating the relationship between slant angle, geometric discretization resolution, and error in potential and forces. 
Finally, section~\ref{sec:compute-time} benchmarks computational runtime and accuracy and offers techniques to improve speed while preserving accuracy.
We conclude in section~\ref{sec:conclusion} with a summary of findings and outlook on future work.

\section{Mathematical Formulation and Validation}\label{sec:mathematical-formulation}

\begin{figure}[htbp]
    \centering
    \includegraphics[width=0.95\linewidth]{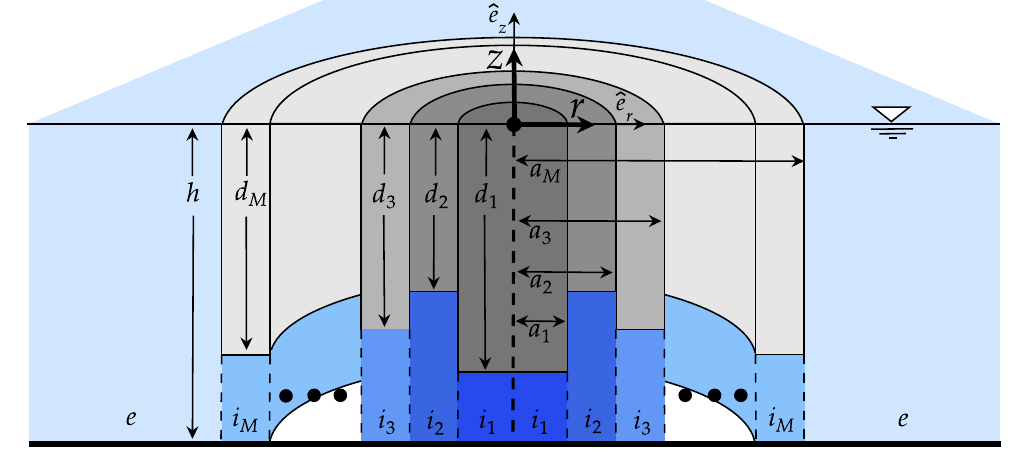}
    \caption{Side view of concentric cylindrical bodies.}
    \label{fig:Diagram}
\end{figure}

The objective of this method is to determine the first-order hydrodynamic forces on a series of $M$ fixed or heaving surface-piercing concentric annular cylinders, 
as shown in Fig.~\ref{fig:Diagram}. 
The internal fluid regions underneath each cylindrical ring are denoted by $i_1$, $i_2$, $\dots, i_M$. 
Each cylindrical ring can be an independent body or rigidly fixed to any other cylindrical ring. 
The external fluid region surrounding the body is denoted by $e$. 
The geometry of the $m$th internal region is defined in terms of the radius $a_m$ measured from the axis of symmetry 
and the draft $d_m$ measured from the mean free surface. 
The origin $O$ is located at the intersection of the body axis of symmetry and the mean free surface of the fluid. 
It is assumed that $a_{m+1}>a_m$ and $d_m<h$ for all $m$, where $h$ is the sea depth. 
This geometry can be used to model more general bodies by discretizing slanted sections into a finite number of annular regions, as we will show in Sec.~\ref{sec:slant}.
Such geometries include the CorPower \citep{de2016techno}, WaveBot \citep{strofer2023control}, AquaHarmonics \citep{weaver2020super}, LUPA float \citep{beringer2025degrees}, and RM3 float \citep{neary2014methodology} WECs.
 
\subsection{Linear Hydrodynamics and Eigenfunctions}
To model the fluid in the internal and external regions, linear potential flow theory is used. 
This implies small body motion, an inviscid, irrotational, and incompressible fluid, and a wave amplitude-to-wavelength ratio less than unity \citep{mei2005theory}. 
With these assumptions, the Navier-Stokes equations simplify to the Laplace equation $\nabla^2 \Phi_\mathrm{T}(\mathbf{x},t)=0$, where the fluid velocity is $\mathbf{v} = \nabla \Phi_\mathrm{T}(\mathbf{x},t)$. 
The potential $\Phi_\mathrm{T}$ is a function of both time $t$ and spatial coordinates $\mathbf{x}$. 
Since this work models only heaving vertically-axisymmetric bodies, the cylindrical coordinate system 
$\mathbf{x}=(r, \theta, z)$ is used, as shown in Fig.~\ref{fig:Diagram}. 
Specifically, the body's radius versus depth profile must be continuous and monotonic,
which excludes cases where the bottom of a fluid region touches a body.
Spatial and temporal dependencies can be separated as 
$\Phi_\mathrm{T}(\mathbf{x},t) = \mathrm{Re} \{ \phi_\mathrm{T}(\mathbf{x})e^{-i \omega t}\}$, 
where $\phi_\mathrm{T} \in \mathbb{C}$ 
and $\omega$ is the angular frequency of the incident wave. 

The total complex potential $\phi_\mathrm{T}$ can be written as 
the superposition of the incident $\phi_\mathrm{I}$, 
diffracted (or scattered) $\phi_\mathrm{D}$, 
and radiated $\phi_\mathrm{R}$ potentials. 
The incident wave potential for a regular wave propagating at an angle $\psi$ from the x-axis is
\begin{equation}
    \phi_\mathrm{I} = -i \frac{g A}{\omega} 
                         \frac{\cosh \lambda_0^e (z+h)}{\cosh \lambda_0^e h}  
                         e^{i\lambda_0^er\cos(\theta- \psi)}
\end{equation}
where $g$ is the acceleration due to gravity, 
$A$ is the wave amplitude, 
$\omega$ is the angular wave frequency, 
$\lambda_0^e$ is the wavenumber, and $h$ is the water depth~\citep{chatzigeorgiou2018analytical}. 
The angle $\psi$ is set to zero without loss of generality, 
as the hydrodynamic forces in heave are invariant with the incident wave direction for vertically axisymmetric bodies. 

The diffracted and radiated wave potentials can be solved by finding velocity potentials that satisfy the boundary conditions on the wetted surface of the body. 
However, to find the first-order hydrodynamic forces on the floating body, only the incident and radiated potential are needed due to the Haskind relation, as explained in Sec.~\ref{Hydrodynamic Forces}. 
Thus, the main focus of Sec.~\ref{Matching Across Fluid Boundaries} and~\ref{Block Matrix Structure} is solving the radiation problem. 
For simplicity, $\phi$ is used throughout the rest of this work 
to denote the complex radiated potential instead of $\phi_\mathrm{R}$. 

To solve the radiation problem, the radiated potential $\phi$ is defined separately 
for the internal and external regions as $\phi^{i_m}$ and $\phi^{e}$, respectively. 
The velocity potentials must satisfy the Laplace equation and boundary conditions 
on the sea floor, wetted surface of the bodies, and free surface. 
The velocity potential in the interior regions can be written as the superposition 
of a homogeneous part $\phi^{i_m}_\mathrm{h}$, which corresponds to the body in region $m$ being fixed, 
and a particular part $\phi^{i_m}_\mathrm{p}$, which corresponds to the body in region $m$ heaving with unit amplitude velocity. 
The total velocity potential in the $m$th region is 
$\phi^{i_m}= \phi^{i_m}_\mathrm{h} + \phi^{i_m}_\mathrm{p}$. 
The velocity potentials can be written as a product of eigenfunctions 
$\phi (r, \theta, z)= R(r)\Theta(\theta)Z(z)$, 
and the Laplace equation can be solved using separation of variables. 
Appendix~\ref{appA} discusses the details of this process.

The corresponding functions for the velocity potential, eigenfunctions, and eigenvalues 
are summarized in Table~\ref{tab:MEEM-eigenfunctions}. 
These solutions are chosen since they satisfy the boundary conditions (Eq.~\ref{Laplace equation internal region}-\ref{eq:R-ODE}) for their corresponding region. 
Note that, when modelling a set of cylindrical rings heaving while all other rings are fixed, 
$\phi^{i_m}_\mathrm{p}$ will only be non-zero in the heaving regions. 
This is indicated by the separate cases of the particular potentials shown in Table~\ref{tab:MEEM-eigenfunctions}. 
$\textrm{I}_0$, $\textrm{K}_0$, and $\textrm{H}_0^1$ 
are the zeroth-order modified Bessel function of the first kind, modified Bessel function of the second kind, and Hankel function of the first kind, respectively. 
All eigenfunctions in Table~\ref{tab:MEEM-eigenfunctions} are real, except $R_{10}^{e}(r)\in\mathbb{C}$.
The expression for $N_{n_e}$ is defined in Appendix~\ref{appA}.
This expression scales the vertical eigenfunctions of the exterior region, and is not to be confused with $N^{e}$ the number of terms in that region. 

As Table~\ref{tab:MEEM-eigenfunctions} shows, after specifying the frequency $\omega$ and geometry of the cylindrical rings, 
the only quantities left unknown are the eigencoefficients $C$ in the series representation of the homogeneous potentials. 
While the exact solutions of the homogeneous potentials are infinite series, in practice the series are truncated as indicated 
by $N^{i_m}$ and $N^{e}$ in Table~\ref{tab:MEEM-eigenfunctions}.  

\begin{landscape}
\begin{table}
    \centering
    \begin{tabular}{|>{\centering\arraybackslash}p{0.085\linewidth}|>{\centering\arraybackslash}p{0.24\linewidth}|>{\centering\arraybackslash}p{0.34\linewidth}|>{\centering\arraybackslash}p{0.24\linewidth}|} \hline 
         Region&  Innermost Interior $(i_1)$&  Interior $(i_m$ for $m>1)$& Exterior $(e)$\\ \hline 
         Homog. potential &  $\phi^{i_1}_\mathrm{h}(r,z) = \displaystyle\sum_{n_1=0}^{N^{i_1}-1} C_{1{n_1}}^{i_1} R_{1{n_1}}^{i_1}(r) Z_{n_1}^{i_1}(z)$&  $\phi^{i_m}_\mathrm{h}(r,z) = \displaystyle\sum_{n_m=0}^{N^{i_m}-1} \left(C_{1{n_m}}^{i_m} R_{1{n_m}}^{i_m}(r) + C_{2{n_m}}^{i_m} R_{2{n_m}}^{i_m}(r) \right) Z_{{n_m}}^{i_m}(z)$& $\phi^{e}_\mathrm{h}(r,z) = \displaystyle\sum_{n_e=0}^{N^{e}-1} C_{1{n_e}}^{e} R_{1{n_e}}^{e}(r) Z_{n_e}^{e}(z)$\\ \hline 
         Partic. potential &  $\phi^{i_1}_\mathrm{p}(r,z) =  \begin{cases} \displaystyle\frac{1}{2(h-d_1)}\left[ (z+h)^2 - \frac{r^2}{2}\right] & \text{Heaving} \\ 0 & \text{Fixed}             
         \end{cases}$&  $\phi^{i_m}_\mathrm{p}(r,z) = \begin{cases} \displaystyle\frac{1}{2(h-d_m)}\left[ (z+h)^2 - \frac{r^2}{2}\right] & \text{Heaving} \\ 0 & \text{Fixed}       
         \end{cases}$& $0$\\ \hline 
         Radial eigenfunction &  $R_{1{n_1}}^{i_1}(r) = \begin{cases}
            \frac{1}{2} &  n_1=0 \\[1em]   
            \frac{\mathrm{I}_0(\lambda_{n_1}^{i_1}r)}{\mathrm{I}_0(\lambda_{n_1}^{i_1}a_1)} & n_1 \ge 1
        \end{cases} $&  \shortstack{$R_{1n_m}^{i_m}(r) = \begin{cases}
            \frac{1}{2} &  n_m=0 \\[1em]   
            \frac{\mathrm{I}_0(\lambda_{n_m}^{i_m}r)}{\mathrm{I}_0(\lambda_{n_m}^{i_m}a_m)} & n_m \ge 1
        \end{cases}$   \\ $R_{2n_m}^{i_m}(r) = \begin{cases}
           \frac{1}{2}\ln(\frac{r}{a_m}) &  n_m = 0 \\
        \frac{\mathrm{K}_0(\lambda_{n_m}^{i_m}r)}{\mathrm{K}_0(\lambda_{n_m}^{i_m}a_m)} & n_m \ge 1
        \end{cases}$}& $R_{1n_e}^{e}(r) = \begin{cases}
           \frac{\mathrm{H}_0^{1}(\lambda_0^e r)}{\mathrm{H}_0^{1}(\lambda_0^e a_M)} & n_e = 0 \\[1em] 
          \frac{\mathrm{K}_0(\lambda_{n_e}^e r)}{\mathrm{K}_0(\lambda_{n_e}^ea_M)} &  n_e \ge 1
        \end{cases}$\\ \hline 
 Vertical eigenfunction & $Z_{n_1}^{i_1}(z) = \begin{cases}
           1 & n_1=0 \\[1em]   
           \sqrt{2}\cos(\lambda_{n_1}^{i_1}(z+h)) & n_1 \ge 1
        \end{cases}$& $Z_{n_m}^{i_m}(z) = \begin{cases}
           1 & n_m=0 \\[1em]   
           \sqrt{2}\cos(\lambda_{n_m}^{i_m}(z+h)) & n_m \ge 1
        \end{cases}$&$    Z^{e}_{n_e}(z) = \begin{cases}
           N_0^{-\frac{1}{2}}\cosh( \lambda_0^{e}(z+h)) &  n_e=0 \\[1em]   
           N_{n_e}^{-\frac{1}{2}}\cos( \lambda_{n_e}^{e}(z+h)) &  n_e \ge 1
        \end{cases}$\\ \hline
 Eigenvalue& $\displaystyle \lambda_{n_1}^{i_1} = \frac{n_1\pi}{h-d_{1}}, \quad  n_1 \geq 1$& $\displaystyle \lambda_{n_m}^{i_m} = \frac{n_m\pi}{h-d_m},  \quad n_m \geq 1$&
 $\displaystyle \begin{cases} \lambda_{0}^{e} \tanh(\lambda_{0}^{e} h)= \omega^2/g, & n_e=0 \\ \lambda_{n_e}^{e} \tan(\lambda_{n_e}^{e} h) = -\omega^2/g, & n_e \geq 1\\ \end{cases} $\\\hline
    \end{tabular}
    \caption{Equations for the potential (homogeneous and particular), eigenfunctions (radial and vertical), and eigenvalues for each region. $i$ and $e$ denote internal and external regions, respectively.}
    \label{tab:MEEM-eigenfunctions}

\end{table}
\end{landscape}

\subsection{Matching Across Fluid Boundaries }\label{Matching Across Fluid Boundaries}

The eigencoefficients must be selected to enforce zero radial fluid velocity at the body boundary and the matching of the potentials and radial velocities at the edges of each region, 
earning this technique the name Matched Eigenfunction Expansion Method (MEEM). 
As shown by the homogeneous and particular potential functions in Table.~\ref{tab:MEEM-eigenfunctions}, 
the radiated potential is a sum of products of eigenfunctions with unknown eigencoefficients 
$C_{1n_m}^{i_m}$, $C_{2n_m}^{i_m}$, and $C_{1n_e}^{e}$. 
Note that $C_{2n_1}^{i_1}=C_{2n_e}^{e}=0$. 
In practice, the infinite series for the $i_1$, $i_m$ for $m>1$, and $e$ regions 
are truncated at $N^{i_1}-1$, $N^{i_m}-1$, and $N^e-1$, yielding a total of 
$N_\mathrm{T} =N^{i_1}+\sum_{m=2}^{M}2N^{i_m}+N^e$ unknown eigencoefficients. 
To solve for these unknowns, radial boundary conditions must be imposed at each interface between regions.

\begin{figure}[h]
    \centering
    \includegraphics[width=0.5\linewidth]{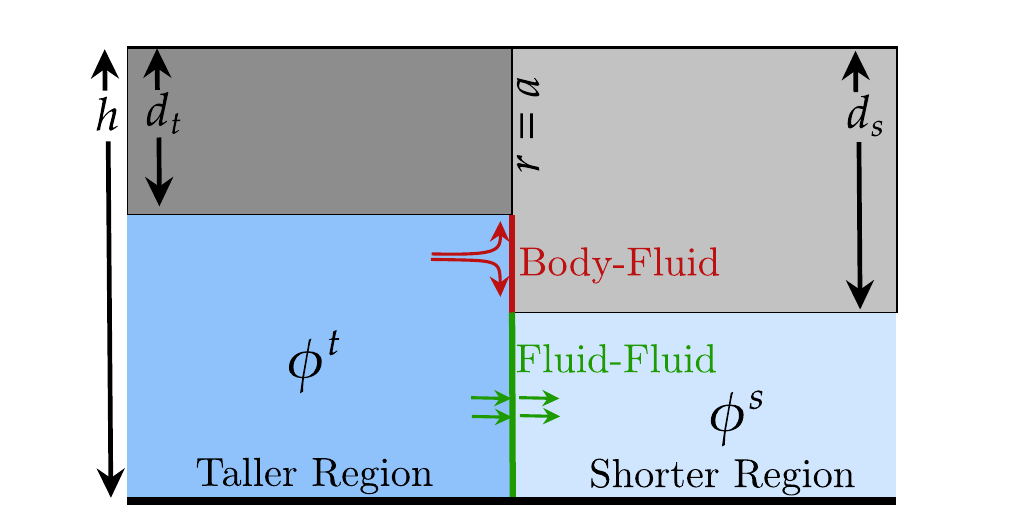}
    \caption{Side view of taller and shorter region.}
    \label{fig:Matching Diagram}
\end{figure}

For a geometry with $M$ internal regions (and one external region), there are $M$ vertical boundaries. 
The general formulation of the matching equations can be illustrated by considering two neighboring fluid regions, 
as shown in Fig.~\ref{fig:Matching Diagram}. 
The two fluid regions will be defined as the taller $\mathrm{t}$ and shorter $\mathrm{s}$ fluid regions. 
When considering the vertical boundary dividing the two regions at $r=a$, 
there are three different conditions to enforce: 
1) the value of the velocity potentials at fluid-fluid boundaries are equal
\begin{equation}\label{potential matching}
    \phi^\mathrm{t}(a,z)=\phi^\mathrm{s}(a,z) 
    \text{ for } -h \le z \le -d_\mathrm{s},
\end{equation}
2) the radial fluid velocities at the fluid-fluid boundary are equal
\begin{equation}\label{velocity matching}
    \frac{\partial \phi^\mathrm{s}}{\partial r}(a,z)
    =\frac{\partial \phi^\mathrm{t}}{\partial r}(a,z) 
    \text{ for } -h \le z \le -d_\mathrm{s},
\end{equation}
and 3) the radial fluid velocity at the body-fluid boundary is zero
\begin{equation}\label{no radial velocity}
    \frac{\partial \phi^\mathrm{t}}{\partial r}(a,z)=0 
    \text{ for } -d_\mathrm{s} \le z \le -d_\mathrm{t}.
\end{equation}
\noindent Using Eq.~\ref{potential matching}-\ref{no radial velocity}, 
the eigenfunctions in Table~\ref{tab:MEEM-eigenfunctions}, 
and the properties of orthogonality, a system of linear algebraic equations can be derived in terms of the eigencoefficients. 
The details of this process are discussed in Appendix~\ref{appB}.

\subsection{Block Matrix Structure}\label{Block Matrix Structure}
The system of equations is in the form $\mathbf{A} \vec{x}=\vec{b}$. 
$\mathbf{A} \in \mathbb{C}^{N_\mathrm{T}\times N_\mathrm{T}}$ 
is a block bi-diagonal matrix composed of sub-matrices 
$\mathbf{A}_1,\mathbf{A}_2,\dots, \mathbf{A}_M$ and zero matrices, 
as shown in Table~\ref{tab:MEEM-A-matrix}. 
The vector of unknowns has the form 
$\vec{x}=[\vec{C}_{1}^{i_1}, \vec{C}_{1}^{i_2}, \vec{C}_{2}^{i_2},\dots, \vec{C}_{1}^{i_M}, \vec{C}_{2}^{i_M}, \vec{C}_{1}^{e}]^T \in \mathbb{C}^{N_\mathrm{T}}$, 
where $\vec{C}_{\chi}^{j} = [C_{\chi1}^j, C_{\chi2}^j, ..., C_{\chi(N^j-1)}^j]$.
Index $j\in\{i_1,i_2,...,i_M,e\}$ specifies the region and $\chi\in\{1,2\}$ specifies the eigenfunction type. 
Column vector $\vec{b} \in \mathbb{R}^{N_\mathrm{T}}$ consists of sub-vectors $\vec{b}_m$ such that 
$\vec{b}=[\vec{b}_1,\vec{b}_2,\dots, \vec{b}_M]^T$. 
Each sub-matrix $\mathbf{A}_m$ and sub-vector $\vec{b}_m$ contains information about the matching of the potential and velocity at $r=a_m$. 

Depending on which fluid region is taller, there are two cases of $\mathbf{A}_m$: 
1) when $d_m>d_{m+1}$, which is shown in Table~\ref{tab:MEEM-A_m-matrix-case-1},  and 
2) when $d_m<d_{m+1}$, which is shown in Table~\ref{tab:MEEM-A_m-matrix-case-2}.
The first case is more common, although the second case can occur for non-convex bodies.
Corresponding right-hand-side vectors $\vec{b}_m$ are shown in 
Table~\ref{tab:MEEM-b_m-vector-case-1} and~\ref{tab:MEEM-b_m-vector-case-2}, respectively. 
The matrix $\boldsymbol{\mathcal{Z}}^{ij}$ has size $N^i \times N^j$ 
and contains the integrals of the products of vertical eigenfunctions in regions $i$ and $j$. 
Closed form expressions for $\boldsymbol{\mathcal{Z}}^{ij}$ are shown in Appendix~\ref{appB}. 
The vector $\vec{R}_\chi^j(r) = [R_{\chi0}^j(r), R_{\chi1}^j(r),\dots, R_{\chi(N^j-1)}^j(r) ]$ is evaluated at $r=a_m$ in each sub-matrix $\mathbf{A}_m$. 
Furthermore, $\mathbf{1}_{ij}$ is a matrix with $i$ rows and $j$ columns containing only ones, 
and $\odot$ is the Hadamard (element-wise) product. 
The equations in Tables~\ref{tab:MEEM-A_m-matrix-case-1}-\ref{tab:MEEM-A_m-matrix-case-2} are valid as-is when $2 \le m \le M-1$. 
When $m=1$, the $\vec{C}_2^{i_1}$ column should be excluded from $\mathbf{A}_1$ 
since the only unknowns associated with $\phi^{i_1}$ are $\vec{C}_1^{i_1}$.
Similarly, when $m=M$, all $i_{m+1}$ indices should be replaced with $e$, 
$d_{m+1}$ should be set to zero, and the $\vec{C}_2^{i_{m+1}}$ column should be excluded from $\mathbf{A}_M$.
Consequently, $\mathbf{A}_m \in \mathbb{R}^{(N^{i_m} + N^{i_{m+1}}) \times (2N^{i_m} + 2N^{i_{m+1}})}$, 
while $\mathbf{A}_1 \in \mathbb{R}^{(N^{i_1} + N^{i_{2}}) \times (N^{i_1} + 2N^{i_2})}$ 
and $\mathbf{A}_M \in \mathbb{C}^{(N^{i_M} + N^{e}) \times (2N^{i_M} + N^e)}$.
This is consistent with Table~\ref{tab:MEEM-A-matrix}. 
Comparing Tables~\ref{tab:MEEM-A_m-matrix-case-1} and \ref{tab:MEEM-A_m-matrix-case-2}, each $\mathbf{A}_m$ sub-matrix contains four square diagonal blocks and four rectangular dense blocks but differs in the location of each block type.
Diagonal blocks arise from orthogonality of the vertical eigenfunctions within a given region,
and dense blocks arise from coupling of the vertical eigenfunctions between adjacent regions.

Tables~\ref{tab:MEEM-b_m-vector-case-1} and \ref{tab:MEEM-b_m-vector-case-2} 
show the two cases of $\vec{b}_m$ sub-vectors, which have closed form solutions.
For $1 \le m \le M-1$, these can be used without modification to obtain all vectors $\vec{b}_m$.
For $m=M$, Table~\ref{tab:MEEM-b_m-vector-case-1} is modified so that any $i_{m+1}$ indices are replaced with $e$ and $d_{m+1}$ is set to zero. 
Additionally, Appendix~\ref{sec:Forms of Matrix A} shows that each sub-matrix $\mathbf{A}_m$ decomposes into an element-wise product of purely radial and purely vertical components: $\mathbf{A}_m = \mathbf{A}_{m,r} \odot \mathbf{A}_{m,z}$.

After constructing the $\mathbf{A}$ matrix and $\vec{b}$ vector 
that correspond to the problem's geometry and wave conditions, 
numerically solving the $\mathbf{A} \vec{x}=\vec{b}$ 
equation yields the unknown eigencoefficients $\vec{x}$. Substituting into the expressions of Table~\ref{tab:MEEM-eigenfunctions} 
yields the velocity potential everywhere in the fluid.
Potential can then be integrated to find the hydrodynamic forces.
\begin{table}
    \centering
    \begin{tabular}{|>{\centering\arraybackslash}p{0.18\linewidth}|c||c|c|c|c|c|c|c|c|c|c|c|c|}
    \hline
     & & $\vec{C}_{1}^{i_1}$& $\vec{C}_{1}^{i_2}$& $\vec{C}_{2}^{i_2}$ & $\vec{C}_{1}^{i_3}$& $\vec{C}_{2}^{i_3}$ & ... 
     &$\vec{C}_{1}^{i_{M-1}}$& $\vec{C}_{2}^{i_{M-1}}$ &$\vec{C}_{1}^{i_M}$& $\vec{C}_{2}^{i_M}$ & $\vec{C}_1^e$ \\\hline 
      &size&  $N^{i_1}$&  $N^{i_2}$&  $N^{i_2}$& $N^{i_3}$&  $N^{i_3}$ &... & $N^{i_{M-1}}$&  $N^{i_{M-1}}$ & $N^{i_M}$ & $N^{i_M}$ & $N^e$\\ \hline \hline 
      
      Boundary 1 & $N^{i_1}+N^{i_2}$ 
        & \multicolumn{3}{c|}{$\mathbf{A}_1$} &\multicolumn{8}{c|}{$\mathbf{0}$} \\ \hline
      
      Boundary 2 & $N^{i_2}+N^{i_3}$
        & $\mathbf{0}$ & \multicolumn{4}{c|}{$\mathbf{A}_2$} &\multicolumn{6}{c|}{$\mathbf{0}$} \\ \hline

      $\vdots$ & $\vdots$
        & \multicolumn{11}{c|}{$\ddots$} \\ \hline

      Boundary $M-1$& $N^{i_{M-1}}+N^{i_M}$
        & \multicolumn{6}{c|}{$\mathbf{0}$} & \multicolumn{4}{c|}{$\mathbf{A}_{M-1}$} & $\mathbf{0}$ \\ \hline

      Boundary $M$ & $N^{i_M}+N^{i_e}$
        & \multicolumn{8}{c|}{$\mathbf{0}$} & \multicolumn{3}{c|}{$\mathbf{A}_M$} \\ \hline
    \end{tabular}
    \vspace{-10pt}
    \caption{MEEM $\mathbf{A}$ matrix.}
    \label{tab:MEEM-A-matrix}
\end{table}

\begin{landscape}
\begin{table}
    \centering
    \begin{tabular}{|>{\centering\arraybackslash}p{0.18\linewidth}|c||c|c|c|c|c|}
    \hline
     & & $\vec{C}_{1}^{i_m}$& $\vec{C}_{2}^{i_m}$& $\vec{C}_{1}^{i_{m+1}}$ & $\vec{C}_{2}^{i_{m+1}}$ \\\hline 
      &size&  $N^{i_m}$&  $N^{i_m}$&  $N^{i_{m+1}}$& $N^{i_{m+1}}$\\ \hline \hline 
      
      \shortstack{
        $\phi^{i_m}=\phi^{i_{m+1}}$ 
        \\ 
        at $r=a_m$} 
        & 
        $N^{i_m}$ 
        & 
        $(h-d_m) \mathrm{diag}\left( \vec{R}_1^{i_m}\right)$ 
        & 
        $(h-d_m) \mathrm{diag}\left( \vec{R}_2^{i_m}\right)$ 
        & 
        $-\boldsymbol{\mathcal{Z}}^{i_mi_{m+1}} \odot \mathbf{1}_{N^{i_m}1} \vec{R}_1^{i_{m+1}}$ 
        & 
        $-\boldsymbol{\mathcal{Z}}^{i_mi_{m+1}} \odot \mathbf{1}_{N^{i_m}1} \vec{R}_2^{i_{m+1}}$ 
        \\ \hline
      
      \shortstack{
        $\frac{\partial}{\partial r}\phi^{i_m}=\frac{\partial}{\partial r}\phi^{i_{m+1}}$ 
        \\ 
        at $r=a_m$} & $N^{i_{m+1}}$
        & 
        $\boldsymbol{\mathcal{Z}}^{i_{m+1}i_m} \odot \mathbf{1}_{N^{i_{m+1}}1} \vec{R}_1^{i_{m}}$ 
        & 
        $\boldsymbol{\mathcal{Z}}^{i_{m+1}i_m} \odot \mathbf{1}_{N^{i_{m+1}}1} \vec{R}_2^{i_{m}}$ 
        & 
        $-(h-d_{m+1}) \mathrm{diag}\left( \frac{\partial}{\partial r} \vec{R}_1^{i_{m+1}}\right)$ 
        & 
        $-(h-d_{m+1}) \mathrm{diag}\left( \frac{\partial}{\partial r} \vec{R}_2^{i_{m+1}}\right)$ 
        \\ \hline
    \end{tabular}
    \caption{MEEM $\mathbf{A}_m$ sub-matrix when $d_m>d_{m+1}$ ($i_m = \mathrm{s}$ and $i_{m+1} = \mathrm{t}$). 
    Note all radial eigenfunctions and their derivatives are evaluated at $r=a_m$.}
    \label{tab:MEEM-A_m-matrix-case-1}
\end{table}
\end{landscape}

\begin{landscape}
\begin{table}
    \centering
    \begin{tabular}{|>{\centering\arraybackslash}p{0.18\linewidth}|c||c|c|c|c|c|}
    \hline
    & & 
    $\vec{C}_{1}^{i_m}$
    & 
    $\vec{C}_{2}^{i_m}$
    & 
    $\vec{C}_{1}^{i_{m+1}}$ 
    & 
    $\vec{C}_{2}^{i_{m+1}}$ 
    \\\hline 
    &
    size
    &  
    $N^{i_m}$
    &  
    $N^{i_m}$
    &  
    $N^{i_{m+1}}$
    & 
    $N^{i_{m+1}}$
    \\ \hline \hline 
      
      \shortstack{
        $\phi^{i_m}=\phi^{i_{m+1}}$ 
        \\ 
        at $r=a_m$} 
        & 
        $N^{i_{m+1}}$ 
        & 
        $-\boldsymbol{\mathcal{Z}}^{i_{m+1}i_m} \odot \mathbf{1}_{N^{i_{m+1}}1} \vec{R}_1^{i_{m}}$ 
        & 
        $-\boldsymbol{\mathcal{Z}}^{i_{m+1}i_m} \odot \mathbf{1}_{N^{i_{m+1}}1} \vec{R}_2^{i_{m}}$ 
        & 
        $(h-d_{m+1}) \mathrm{diag}\left( \vec{R}_1^{i_{m+1}}\right)$ 
        & 
        $(h-d_{m+1}) \mathrm{diag}\left( \vec{R}_2^{i_{m+1}}\right)$ 
        \\ \hline
      
      \shortstack{
        $\frac{\partial}{\partial r}\phi^{i_m}=\frac{\partial}{\partial r}\phi^{i_{m+1}}$ 
        \\ 
        at $r=a_m$} & $N^{i_{m}}$
        & 
        $-(h-d_{m}) \mathrm{diag}\left( \frac{\partial}{\partial r} \vec{R}_1^{i_{m}}\right)$ 
        & 
        $-(h-d_{m}) \mathrm{diag}\left( \frac{\partial}{\partial r} \vec{R}_2^{i_{m}}\right)$ 
        & 
        $\boldsymbol{\mathcal{Z}}^{i_mi_{m+1}} \odot \mathbf{1}_{N^{i_{m}}1} \vec{R}_1^{i_{m+1}}$ 
        & 
        $\boldsymbol{\mathcal{Z}}^{i_mi_{m+1}} \odot \mathbf{1}_{N^{i_{m}}1} \vec{R}_2^{i_{m+1}}$ 
        \\ \hline
    \end{tabular}
    \caption{MEEM $\mathbf{A}_m$ sub-matrix when $d_m<d_{m+1}$ ($i_m = \mathrm{t}$ and $i_{m+1} = \mathrm{s}$). Note all radial eigenfunctions and their derivatives are evaluated at $r=a_m$.}
    \label{tab:MEEM-A_m-matrix-case-2}
\end{table}
\end{landscape}

\begin{table}
    \centering
    \begin{tabular}{|>{\centering\arraybackslash}p{0.18\linewidth}|c||c|c|}
      \hline
      &size& \\ \hline \hline 
      
      \shortstack{$\phi^{i_m}=\phi^{i_{m+1}}$ \\ at $r=a_m$} & $N^{i_m}$ & $\int_{-h}^{-d_m}\left( \phi_\mathrm{p}^{i_{m+1}} - \phi_\mathrm{p}^{i_m} \right) \vec{Z}^{i_m} \mathrm{d}z$\\ \hline
      
      \shortstack{$\frac{\partial}{\partial r}\phi^{i_m}=\frac{\partial}{\partial r}\phi^{i_{m+1}}$ \\ at $r=a_m$} & $N^{i_{m+1}}$
        & $\int_{-h}^{-d_{m+1}}\frac{\partial \phi_\mathrm{p}^{i_{m+1}}}{\partial r} \vec{Z}^{i_{m+1}} \mathrm{d}z-\int_{-h}^{-d_m}\frac{\partial \phi_\mathrm{p}^{{i_{m}}}}{\partial r} \vec{Z}^{i_{m+1}}{} \mathrm{d}z$ \\ \hline
    \end{tabular}
    \caption{MEEM $\vec{b}_m$ vector when $d_m>d_{m+1}$ ($i_m = \mathrm{s}$ and $i_{m+1} = \mathrm{t}$). Note all radial eigenfunctions and their derivatives are evaluated at $r=a_m$.}
    \label{tab:MEEM-b_m-vector-case-1}
\end{table}

\begin{table}
    \centering
    \begin{tabular}{|>{\centering\arraybackslash}p{0.18\linewidth}|c||c|c|}
    \hline 
      &size& \\ \hline \hline 
      
      \shortstack{$\phi^{i_m}=\phi^{i_{m+1}}$ \\ at $r=a_m$} & $N^{i_{m+1}}$ & $\int_{-h}^{-d_{m+1}}\left( \phi_\mathrm{p}^{i_{m}} - \phi_\mathrm{p}^{i_{m+1}} \right) \vec{Z}^{i_{m+1}} \mathrm{d}z$\\ \hline
      
      \shortstack{$\frac{\partial}{\partial r}\phi^{i_m}=\frac{\partial}{\partial r}\phi^{i_{m+1}}$ \\ at $r=a_m$} & $N^{i_{m}}$
        &  $\int_{-h}^{-d_m}\frac{\partial \phi_\mathrm{p}^{i_{m}}}{\partial r} \vec{Z}^{i_{m}} \mathrm{d}z-\int_{-h}^{-d_{m+1}}\frac{\partial \phi_\mathrm{p}^{i_{m+1}}}{\partial r} \vec{Z}^{{i_{m}}} \mathrm{d}z$\\ \hline
    \end{tabular}
    \caption{MEEM $\vec{b}_m$ vector when $d_m<d_{m+1}$ ($i_{m+1} = \mathrm{s}$ and $i_{m} = \mathrm{t}$). 
    Note all radial eigenfunctions and their derivatives are evaluated at $r=a_m$.}
    \label{tab:MEEM-b_m-vector-case-2}
\end{table}

\subsection{Hydrodynamic and Hydrostatic Forces}\label{Hydrodynamic Forces}
In this section, we will characterize the radiation, excitation, and hydrostatic forces of a system with $M$ internal regions and $Q$ heave degrees of freedom. Also, while there are $M$ internal regions, multiple regions may form a single body if the regions are rigidly fixed to one another. Thus, for a system with a total of $M$ internal regions and $Q$ heave degrees of freedom (DOFs), $Q \le M$. 

Moving forward, $\mathbf{A} \vec{x}_q = \vec{b}^q$ will indicate the system of equations associated with the motion of only the $q$th body (and DOF) of the system, while all other bodies are fixed. Thus, $\vec{b}^q=[\vec{b}^q_1,\vec{b}^q_2,...,\vec{b}^q_M]^T$. Note that the matrix $\mathbf{A}$ does not depend on which body is moving. Meanwhile, $\vec{b}^q$, and hence the solution $\vec{x}_q$, does. This is because $\vec{b}^q$ contains integrals of particular potentials, which are zero for stationary regions and non-zero for moving regions.

First, we will find the total radiation force $\vec{f}_{pq}(t)$ on the $p$th body due to the $q$th DOF. 
This can be written in the frequency domain as 
\begin{equation}\label{eq:freq domain vector of rad force}
    \vec{\hat{f}}_{pq}(\omega) = \text{i} \omega  \rho \iint_{S_p} \ _{}^{q}\phi(r,z) \ \hat{n}_p dS
\end{equation}
where $\rho$ is the density of the fluid, $S_p$ is the wetted surface of body $p$, $\hat{n}_p$ is the unit vector normal to the wetted surface of body $p$ pointing outward from the fluid, $\vec{f}_{pq}(t) = \mathrm{Re} \{ \vec{\hat{f}}_{pq} e^{- \text{i} \omega t} \}$, and $_{}^{q}\Phi(\mathbf{x},t) = \mathrm{Re} \{ _{}^{q}\phi(r,z) e^{- \text{i} \omega t} \}$. The left superscript of $q$ is added to distinguish the velocity potentials for different radiation problems. Taking the dot product of Eq.~\ref{eq:freq domain vector of rad force} with the unit vector $\hat{e}_z$ yields the complex heave force on body $p$ due to the motion of body $q$ 
\begin{equation}\label{eq:freq domain scalar rad force}
    \hat{f}_{pq} = \vec{\hat{f}}_{pq} \cdot \hat{e}_z = \text{i} \omega \rho \iint_{S_p} \ _{}^{q}\phi(r,z) \ (\hat{n}_p \cdot \hat{e}_z ) \  dS.
\end{equation}
Since $\hat{n}_p=\hat{e}_z$ at the horizontal portions of $S_p$ and $\hat{n}_p=\hat{e}_r$ at any vertical portions of $S_p$, integration on only the bottom surface of each region contributes to forces in heave. 
Proceeding with integration along the bottom body boundaries, the total heave force on body $p$ will be due to integrating the potential on the bottom boundaries of all cylindrical rings belonging to body $p$. We will define $\mathcal{M}_p$ as the set of all indices that correspond to regions which form the $p$th body. For example, if body 1 consists of regions $i_1$, $i_3$ and $i_4$, $\mathcal{M}_1 = \{1, 3, 4\}$. Eq.~\ref{eq:freq domain scalar rad force} can be written in terms of the potential at each region by
\begin{equation}\label{eq:freq domain scalar rad force in terms of regions}
    \hat{f}_{pq} = \textrm{i} \omega \rho \sum_{m \in \mathcal{M}_p}\int_0^{2 \pi} \int_{a_m}^{a_{m+1}} \ _{}^{q}\phi^{i_m}(r,-d_m) \ r  \ dr\ d\theta
\end{equation}
where $_{}^{q}\phi^{i_m}(r,-d_m)$ is the potential in internal region $i_m$ evaluated at $z=-d_m$ when only the $q$th body is moving. $\hat{f}_{pq}$ can be rewritten in terms of frequency-dependent added mass and radiation damping coefficients $A_{pq}(\omega)$ and $B_{pq}(\omega)$, respectively, which is shown in Appendix~\ref{appC}. The result is
\begin{equation}\label{A_pq B_pq scalar form}
    A_{pq}(\omega) + \frac{\mathrm{i}B_{pq}(\omega)}{\omega}=2\pi \rho (c_p\delta_{pq} + \vec{c}_p \ \vec{x}_q) =2\pi \rho (c_p\delta_{pq} + \vec{c}_p \ \mathbf{A}^{-1} \vec{b}^q)
\end{equation}
with output scalar $c_{p} \in \mathbb{R}$
\begin{equation}
     c_{p} =  \sum_{m \in \mathcal{M}_p } \frac{\left({a^2_{{m+1}}}-{a^2_{m}}\right)\,\left(-{a^2_{m}}-{a^2_{{m+1}}}+4(h-d_m)^2\right)}{16\,\left(h - d_m\right)}, 
\end{equation}
where $\delta_{pq}$ is the Kronecker delta symbol, and output row vector $\vec{c}_p =[\vec{c}_p^{ \ i_1}, \ \vec{c}_p^{ \ i_2}, \dots,  \vec{c}_p^{ \ i_M},  \vec{c}_p^{ \ e}]\in \mathbb{C}^{N_\mathrm{T}}$, where
\begin{equation}\label{eq:c_q_i1_vec}
    \vec{c}_p^{ \ i_1} = \left\{\begin{matrix} \vec{Z}^{i_1}(-d_1) \odot\vec{\mathcal{R}}_{1}^{i_1} & \text{when} \ m \in \mathcal{M}_p \\
    \vec{0}_{N^{i_1}} & \text{when} \ m \notin \mathcal{M}_p
    \end{matrix}\right. ,
\end{equation}
\begin{equation}\label{eq:c_q_im_vec}
    \vec{c}_p^{ \ i_m} = \left\{\begin{matrix} [\vec{Z}^{i_m}(-d_m) \odot\vec{\mathcal{R}}_{1}^{i_m}, \ \vec{Z}^{i_m}(-d_m) \odot\vec{\mathcal{R}}_{2}^{i_m}] & \text{when} \ m \in \mathcal{M}_p \vspace{0.5em} \\
    
    \vec{0}_{2N^{i_m}} & \text{when} \ m \notin \mathcal{M}_p
    \end{matrix}\right. ,
\end{equation}
for $2\le m \le M$, and $\vec{c}_p^{ \ e} = \vec{0}_{N^{e}}$. 
Note that $\vec{0}_{k}$ denotes a $k$-long row vector of zeros and $\vec{\mathcal{R}}_{\chi}^{j}$ is a vector of integrals of radial eigenfunctions for region $j$ with closed form solutions in Appendix~\ref{appC}. 
The vector $\vec{c}_p$ contains non-zero elements that are positioned so they multiply the corresponding eigencoefficients for the $m$th region in the $\vec{x}_q$ vector in Eq.~\ref{A_pq B_pq scalar form}. 
By applying Eq.~\ref{A_pq B_pq scalar form} for $q=1,2,\dots, Q$ and $p=1,2,\dots, Q$, we can express the added mass matrix $\mathbf{A}_\mathrm{r}(\omega) \in \mathbb{R}^{Q \times Q}$ and radiation damping matrix $\mathbf{B}_\mathrm{r}(\omega) \in \mathbb{R}^{Q \times Q}$ as
\begin{equation}\label{A_pq B_pq matrix form}
    \mathbf{A}_\mathrm{r}(\omega) + \frac{\mathrm{i}\mathbf{B}_\mathrm{r}(\omega)}{\omega}=2\pi \rho (\mathbf{C}_0 + \mathbf{C} \ \mathbf{X}) =2\pi \rho (\mathbf{C}_0 + \mathbf{C} \ \mathbf{A}^{-1} \ \mathbf{B})
\end{equation}
where the element in the $p$th row and column of the diagonal matrix $\mathbf{C}_0$ is $c_{p}$, the $p$th row of $\mathbf{C}$ is $\vec{c}_p$, the $q$th column of $\mathbf{X}$ is $\vec{x}_q$, and the $q$th column of $\mathbf{B}$ is $\vec{b}^q$. Finally, the added mass and radiation damping matrices are 
\begin{equation}\label{eq: added mass and damping matrices}
    \mathbf{A}_\mathrm{r}(\omega) = 2\pi \rho \ \mathrm{Re} \{ \mathbf{C}_0 + \mathbf{C} \ \mathbf{A}^{-1} \ \mathbf{B} \} \quad \text{and} \quad
    \mathbf{B}_\mathrm{r}(\omega) = 2\pi \rho \omega \ \mathrm{Im} \{ \mathbf{C}_0 + \mathbf{C} \ \mathbf{A}^{-1} \ \mathbf{B} \}.
\end{equation}

To find the heave excitation force $X_q$ on the $q$th body due to an incident wave, a form of the Haskind relation can be used \citep{newman2018marine}. This was done in~\citet{chau2012inertia} and~\citet{zhang_performance_2024} for geometries with two and three internal regions, respectively. The results are the same for this configuration, as the derivation for the force on the $q$th body only involves the solution in the external region when the $q$th body is heaving. The heave excitation force on the $q$th body is
\begin{equation}\label{eq:excitation force}
    X_q = \frac{-4 \text{i} \rho g h \sqrt{N_0} }{\cosh(\lambda_0^eh)\text{H}_0^1(\lambda_0^e a_M)} \ _{}^{q}C_{10}^e
\end{equation}
where $g$ is the acceleration due to gravity, $\lambda_0^e$ is the wavenumber, $\textrm{H}_0^1$ is the zeroth-order Hankel function of the first kind, and $_{}^{q}C_{10}^e$ is the eigencoefficient in the external region for $n_e=0$ when only the $q$th body is heaving.
By applying Eq.~\ref{eq:excitation force} for $q=1,2,\dots, Q$, we can find the heave excitation force coefficient vector $\vec{X} \in \mathbb{C}^Q$, which is a column vector with the element $X_q$ in the $q$th entry.


The hydrostatic stiffness matrix $\mathbf{K}$ and mass matrix $\mathbf{M}$ are diagonal matrices that can be found from geometry. The element in the $q$th row and $q$th column of $\mathbf{K}$ can be found by summing over the waterplane areas $W_m$ contributed by each region
\begin{equation}\label{eq:hydrostatic stiffness}
    [\mathbf{K}]_{qq}= \rho g \sum_{m \in \mathcal{M}_q} W_m
\end{equation}
where $W_m= \pi a_m^2$ for $m=1$ and $W_m= \pi(a_m^2-a_{m-1}^2)$ otherwise.
If we assume the configuration in Fig.~\ref{fig:Diagram} is in static equilibrium such that the gravitational force balances the buoyancy force, the element in the $q$th row and $q$th column of $\mathbf{M}$, is
\begin{equation}\label{eq:mass matrix}
    [\mathbf{M}]_{qq}= \rho \sum_{m \in \mathcal{M}_q} W_md_m.
\end{equation}
This is the mass of the $q$th body. Once all matrices are found, one can construct the equation of motion of the system for regular waves
\begin{equation}\label{eq: EOM}
    (\mathbf{M} + \mathbf{A}_\mathrm{r}(\omega)) \ddot{\vec{\xi}} + \mathbf{B}_\mathrm{r}(\omega) \dot{\vec{\xi}} + \mathbf{K} \vec{\xi} =\mathrm{Re} \{ A \vec{X} e^{-\mathrm{i} \omega t} \}
\end{equation}
where $\vec{\xi}(t) = [\xi_1(t), \xi_2(t), \dots,  \xi_Q(t)]^T$ is a vector containing the heave displacements of the $Q$ bodies of the system, $\omega$ is the wave frequency, and $A$ is the wave amplitude.

\subsection{Low, High, and Infinite Frequency Approximations}
The wave frequency $\omega$ and wavenumber $\lambda_0^e$ are related by the dispersion relation in Table \ref{tab:MEEM-eigenfunctions}, which is monotonic and depends on $h$. Extreme values of $\lambda_0^e$ and/or $h$ push solutions towards edge cases, enabling simplification or requiring modifications to avoid numerical errors.

\subsubsection{Low Frequency}\label{sec:low-freq}
\begin{figure}[htbp]
    \centering    \includegraphics[width=\linewidth]{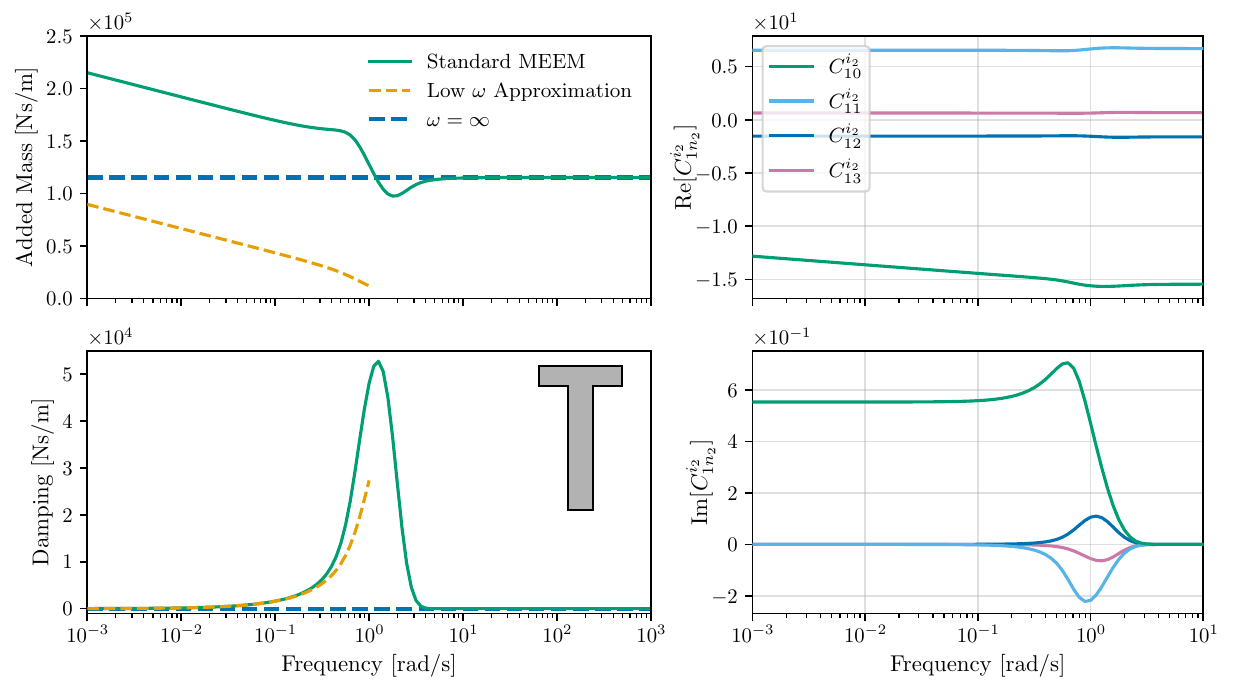}
    \caption{Left: Comparison of the low frequency approximation, the infinite frequency limit, and standard MEEM (with $N^{i_m} = N^e = 100$) for the geometry described in Section~\ref{sec:validation}. Right: A comparison of the first four $C^{i_2}_{1n}$ over low frequencies, demonstrating that the behaviour of $C^{i_2}_{10}$ eventually dominates as frequency approaches zero for both the real and imaginary parts. This trend is representative of the other $C^{i_m}_{1n}$ and $C^{i_m}_{2n}$. Figure context: \href{https://calkit.io/symbiotic-engineering/openflash/figures?path=pubs\%2FJFM\%2Ffigs\%2FMEEM-Low-Freq.pdf}{Calkit} and \href{https://cocalc.com/rmccabe/MEEM/meem}{CoCalc}.}
    \label{fig: extreme-frequencies}
\end{figure}

As $\omega \to 0$, the components of $\vec x$ exhibit the following asymptotic behaviour: $\text{Re}(C^{i_m}_{1n_m})$ and $\text{Re}(C^{i_m}_{2n_m})$ behave like $K_1 + K_2\ln(\lambda_0^e h)$ for $n_m = 0$, where $K_1$ and $K_2$ are nonzero constants that are different for each coefficient and region, and approach nonzero constants for $n_m>0$. Meanwhile, $\text{Im}(C^{i_m}_{1n_m})$ and $\text{Im}(C^{i_m}_{2n_m})$ approach nonzero constants for $n_m=0$ and zero for $n_m>0$, as shown in Fig.~\ref{fig: extreme-frequencies}.

Since $\vec c_p$ is frequency independent, the hydrodynamic coefficients vary with frequency only through their dependence on $C^{i_m}_{1n_m}$ and $C^{i_m}_{2n_m}$.
From their forms, we can see that as $\omega \to 0$, the behaviors of the $n_m = 0$ coefficients dominate the others. Consequently, $A_{11}(\omega)$ grows proportional to $\ln(\lambda_0^e h)$ (although the constant term is still significant for the frequency range in Fig.~\ref{fig: extreme-frequencies}) while $B_{11}(\omega)/\omega$ approaches a constant.
This is consistent with behaviour of the analytical low frequency limits for a single heaving cylinder described in \citet{yeung_added_1981}, $\lim_{\omega \to 0} A_{11}(\omega) = K - \frac{2B_{11}(0)}{\pi\omega}\ln(\lambda_0^e h) \approx - \frac{2B_{11}(0)}{\pi\omega}\ln(\lambda_0^e h)$, where $K$ is a constant. 


The low frequency approximation uses the shallow water approximation to represent the potential in each region with a depth-averaged potential \citep{chau2012inertia}, removing the potential's $z$-dependence. For the body regions, this is equivalent to keeping only the $n_m=0$ eigenfunctions and coefficients, since $Z^{i_m}_0(z) = 1$. For the exterior region, the only characteristic lengths are $1/\lambda_0^e$ and $h$, so low frequency is equivalent to shallow water. However, as the ratio of body radius to water depth ($a_M/h$) increases, the approximation becomes less accurate \citep{yeung_added_1981}. This can be interpreted as the $n_m = 0$ terms dominating faster for shallower water.

\subsubsection{High Frequency}\label{sec:high-freq-limit}
The $\sinh$ component of $N_0$ (and therefore $N_0$) increases exponentially with high $\lambda_0^e h$. $N_0$ appears in the denominator of the first exterior region vertical eigenfunction $Z_0^e$ and its derivative, and anywhere else it appears is a specific case of one of these expressions. Both have hyperbolic functions of $\lambda_0^e(z+h)$ in the numerator that overflow and raise errors long before the fraction as a whole becomes so extreme. We found the expressions' limiting forms to extend their allowed input range. The accuracy of these forms depends solely on the product $\lambda_0^e h$, not $\lambda_0^e $ or $h$ individually, or $z$.

\begin{equation}
	\lim_{\lambda_0^e h \to \infty} Z_0^e(z) = \lim_{\lambda_0^e h \to \infty} \frac{\cosh(\lambda_0^e(z + h))}{\sqrt{N_0}} = 
\sqrt{2 \lambda_0^e h} \left(e^{\lambda_0^e z} + e^ {-\lambda_0^e(z + 2h)}\right)
\end{equation}
\begin{equation}
	\lim_{\lambda_0^e h \to \infty} \frac{\partial Z_0^e(z)}{\partial z} = \lim_{\lambda_0^e h \to \infty} \frac{\lambda_0^e\sinh(\lambda_0^e(z + h))}{\sqrt{N_0}} = \lambda_0^e
\sqrt{2 \lambda_0^e h}  \left(e^{\lambda_0^e z} - e^ {-\lambda_0^e(z + 2h)}\right)
\end{equation}
Empirically, the approximated expressions are less than a fraction of $10^{-10}$ off from their true values for $\lambda_0^e h > 14$. This was encoded as the threshold for using the approximations.

\subsubsection{Infinite Frequency}\label{sec:inf-frequency}
As $\lambda_0^e$ increases, the coefficient of the first exterior region eigenfunction decreases, and later eigenfunctions dominate. At $\lambda_0^e = \infty$,
the solution is representable without the first exterior region eigenfunction. The rest of the exterior region eigenvalues must be finite and satisfy $\lambda_n^e \tan (\lambda_n^e h) = - \infty$, meaning $\lambda_n^e h = (n - \frac{1}{2})\pi$ and
\begin{equation} \lim_{\lambda_0^e \to \infty }\lambda_n^e = \frac{(n - \frac{1}{2})\pi}{h}.
\end{equation}
In general, that is the lower bound for $\lambda_n^e$. 

Lastly, damping approaches zero as $\lambda_0^e$ approaches infinity. Mathematically, this is evident from the matrix formulation: the Hankel functions $H_0^1$ are the only Bessel functions involved that give imaginary values for real inputs, so they supply the only imaginary elements to the $\mathbf{A}$ matrix. When their contribution goes to zero, the matrix (and its solution) become real, leaving no imaginary component in the hydrodynamic coefficient integral.

\subsection{Numerics}\label{sec:numerics}
This section discusses numerical considerations including overflow, finite precision, condition number, and numerical solution algorithms that become relevant when implementing MEEM. 
Avoiding numerical failures in edge cases is particularly relevant for optimization applications where unusual geometries may be evaluated at intermediate iterations. In the following subsections, we denote the maximum representable float before overflow as $\rm{f}_{max}$.


\subsubsection{Overflow in Radial Eigenfunctions}
Without using exponentially scaled Bessel functions (Table~\ref{table:exp-bessels}), the radial eigenfunctions overflow at arguments near $\lambda r \text{ (and/or $\lambda a$)}\geq c_\mathrm{overflow} \approx \ln(\rm{f}_{max})$ due to overflow in the numerator or denominator individually, where $\lambda$ can be $\lambda_{n_e}^e$ or $\lambda_{n_m}^{i_m}$. 
Meanwhile, the radial eigenfunctions computed using exponentially scaled Bessel functions will only overflow after $|\lambda (r - a)|\geq c_\mathrm{overflow}$.
This allows larger values of $\lambda a$ as long as $r$ is near $a$, extending the dimensions of geometries and terms per region allowed.
Due to the relationship for $\lambda_{n_m}^{i_m}$ in Table~\ref{tab:MEEM-eigenfunctions}, the use of the exponentially scaled Bessel functions alleviates the restriction on the maximum allowable truncation order of the series in region $i_m$ from $N^{i_m} < c_\mathrm{overflow}(h-d_m) / (\pi a_m)$ to $N^{i_m} < c_\mathrm{overflow}(h-d_m) / (\pi(a_{m+1} - a_m))$. Notice how this new restriction depends on the difference in radial dimensions rather than the radial dimension itself. Since this permits the use of larger truncation orders, more accurate solutions are achievable.

\begin{table}
\centering \begin{tabular}{|c|c|c|}
\hline
Exponential Scaling Formula & Replacement Example\\
\hline
$I_\nu^e(z) = I_\nu(z) \cdot e^{-z}$ & 
$\frac{I_{\nu}(\lambda r)}{I_\nu(\lambda a)} = \frac{I_\nu^e (\lambda r)}{I_\nu^ e(\lambda a)}\cdot e^{\lambda(r-a)}$\\
\hline
$K_\nu^e(z) = K_\nu(z) \cdot e^{z}$ &
$\frac{K_{\nu}(\lambda r)}{K_\nu(\lambda a)} = \frac{K_\nu^e (\lambda r)}{K_\nu^ e(\lambda a)}\cdot e^{\lambda(a-r)}$\\
\hline
\end{tabular}
\caption{Typical Bessel functions ($I_\nu(z), K_\nu(z)$) exhibit approximately exponential growth or decay. Exponentially scaled Bessel functions counteract this and span only a few orders of magnitude for the reasonable range of inputs. Examples of the new form of the radial eigenfunctions are shown here.} \label{table:exp-bessels} \end{table}

\subsubsection{Overflow in Vertical Eigenfunctions}
The vertical eigenfunction $Z_k^e$ for $k=0$ contains the $\cosh$ and $\sinh$ functions, which diverge for large values of $\lambda_0^eh$ (high frequencies or deep water).
The largest relevant value of $\lambda_0^eh$ depends on the environment rather than on the floating body geometry.
The expression overflows for $\lambda_0^eh>\cosh^{-1}(\rm f_{max})/2$.
To prevent overflow, we use the analytical limit derived in section~\ref{sec:high-freq-limit}, which allows for larger values of $\lambda_0^eh>\cosh^{-1}(\rm f_{max})h/d_2$ before overflowing.
Substituting this into the first element of the bottom block of the b-vector results in 
\begin{equation}
    \lim_{\lambda_0^eh\rightarrow\infty}[\vec{b}_M]_0=\frac{-a_M}{h-d_M} \sqrt{\frac{h}{2\lambda_0^e}}\exp(-d_M\lambda_0^e)
\end{equation}



and the corresponding limit for the vertical coupling integral
\begin{equation}
\begin{aligned}
    \lim_{\lambda_0^eh\rightarrow\infty}[\boldsymbol{\mathcal{Z}}^{i_m e}]_{n_m 0} &
    = h~\frac{\cosh^2(\lambda_0^eh)}{\sqrt{2\lambda_0^eh}}\cdot \frac{-1+(-1)^{n_m}\exp(1-\frac{d_M}{h})}{f_{n_m}} \\
    \text{where}~~f_{n_m}&= \begin{cases}
        1, & n_m=0 \\
        h^2{(\lambda^{i_m}_{n_m})}^2+1, & n_m \geq 1
    \end{cases}
    \end{aligned}.
\end{equation}

\subsubsection{Nonlinear Solve of $\lambda_{n_e}^e$}\label{sec:nonlin-solve}
A final numerical subtlety worth discussing is finite precision effects in calculating $\lambda_{n_e}^e$.
Bounds of $180^\textrm{o}\cdot[n_e-\frac{1}{2}, n_e]$ are placed on $\lambda_{n_e}^eh$ in a root-finding algorithm to ensure the $n_e$th root is identified.
Degrees are used instead of radians so asymptotes occur at rational values.

\subsubsection{Matrix Condition Number}
The $\mathbf{A}_m$ sub-matrix, and consequently the $\mathbf{A}$ matrix, frequently has a high condition number. 
This arises because of the large range of the radial eigenfunction across values of $n_m$.
More specifically, using the exponentially scaled modified bessel function of the first kind from Table~\ref{table:exp-bessels}, 
the ratio of the $n_m=1$ term to the $n_m=N^{i_m}-1$ term of the $\vec{R}_{1}^{i_m}$ vector in a given block of the $\mathbf{A}_m$ sub-matrix scales approximately with $\exp(2\pi (N^{i_m}-2) a_m/(h-d_m))$.
This means that the condition number of $\mathbf{A}$ grows roughly exponentially with the maximum truncation order of the series, and more rapidly for geometries in which large radii co-occur with large drafts.
In principle, a high condition number means that the $\mathbf{A}$ matrix will amplify small deviations or numerical errors in the $\vec{b}$ vector to produce large errors in the $\vec{x}$ vector, or equivalently that the $\vec{x}$ vector is close to non-unique.
However, the validation in section~\ref{sec:validation} will show that these high condition numbers do not appear to interfere with the accuracy or convergence of the solution, even when the linear algebra solver produces warnings about ill-conditioned inputs.
In their review of a related method, \cite[Section 4.3]{hiptmair_survey_2016} explain the high condition number as a consequence of the eigenfunctions becoming nearly linearly dependent at large truncation orders.
They reference some studies which likewise find no issue with the high condition number and others which choose the number of terms per region to balance truncation error and conditioning.

\subsubsection{Matrix Sparsity}
\begin{figure}[h!]
    \centering
    \begin{subfigure}{0.58\textwidth}
        \includegraphics[width=\textwidth]{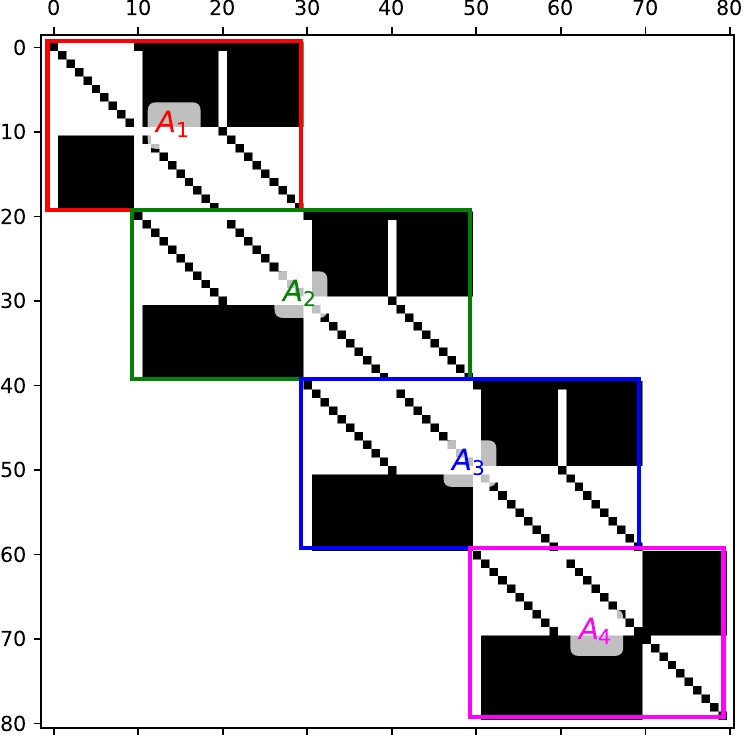}
        \caption{$\mathbf{A}$}
        \label{fig:sparsityA}
    \end{subfigure}
    \hfill
    \begin{subfigure}{0.09\textwidth}
        \includegraphics[width=\textwidth]{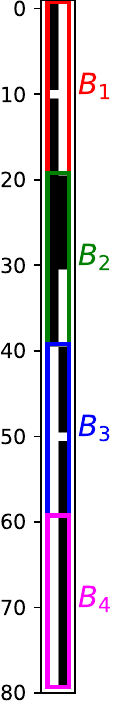}
        \caption{$\mathbf{B}$}
        \label{fig:sparsityB}
    \end{subfigure}
    \hfill
    \begin{subfigure}{0.09\textwidth}
        \includegraphics[width=\textwidth]{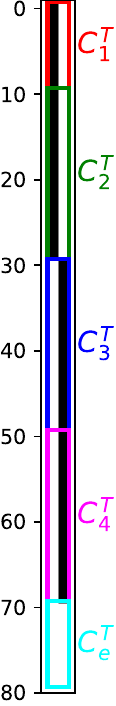}
        \caption{$\mathbf{C}^T$}
        \label{fig:sparsityC}
    \end{subfigure}
    \caption{Sparsity pattern in the $\mathbf{A}$, $\mathbf{B}$, and $\mathbf{C}$ matrices. Figure context: \href{https://calkit.io/symbiotic-engineering/openflash/notebooks?path=analysis\%2Fsparsity_plots.ipynb}{Calkit} and \href{https://cocalc.com/rmccabe/MEEM/meem}{CoCalc}.}
    \label{fig:sparsity}
\end{figure}

Fig.~\ref{fig:sparsity} shows the sparsity pattern of the $\mathbf{A}$, $\mathbf{B}$, and $\mathbf{C}$ matrices from Eq.~\ref{eq: added mass and damping matrices} for an example configuration consisting of two bi-cylinder bodies with $M=4$, $\mathcal{M}_1 = \{1, 2\}$, $\mathcal{M}_2 = \{3, 4\}$, $N^{i_m}=N^e=10$, and $d_{i+1}<d_i$. 
The $\mathbf{A}$ matrix features a block bi-diagonal structure.
This can be observed by partitioning $\mathbf{A}$ into rectangular blocks such that each block encompasses the full height and half the blockwise width of the $\mathbf{A}_m$ sub-matrix.
The bi-diagonal structure occurs because the matching equations at a given boundary contain only unknowns from the two regions at that boundary, and not from other regions.
Each block features its own sparsity pattern, where some portions are completely populated, 
and others have multiple individual bands of elements due to the diagonal blocks in Tables~\ref{tab:MEEM-A_m-matrix-case-1} and \ref{tab:MEEM-A_m-matrix-case-2}. 
Although a standard linear solver was used in this work (\texttt{scipy}'s LU decomposition), 
the computational complexity of MEEM may be reduced further by exploiting the sparsity of $\mathbf{A}$. 
This is discussed further in Section~\ref{sec:compute-time}.

The $\mathbf{B}$ matrix features a sparsity pattern that is dependent on the degrees of freedom prescribed to each cylindrical ring. 
Finally, the sparsity pattern in the $\mathbf{C}$ matrix depends on which cylindrical rings are influenced. 
This is a consequence of Eq.~\ref{eq:c_q_i1_vec} and \ref{eq:c_q_im_vec}.

\subsection{Validation}\label{sec:validation}
\begin{figure}[htbp]
    \centering
    \includegraphics[width=0.95\linewidth]{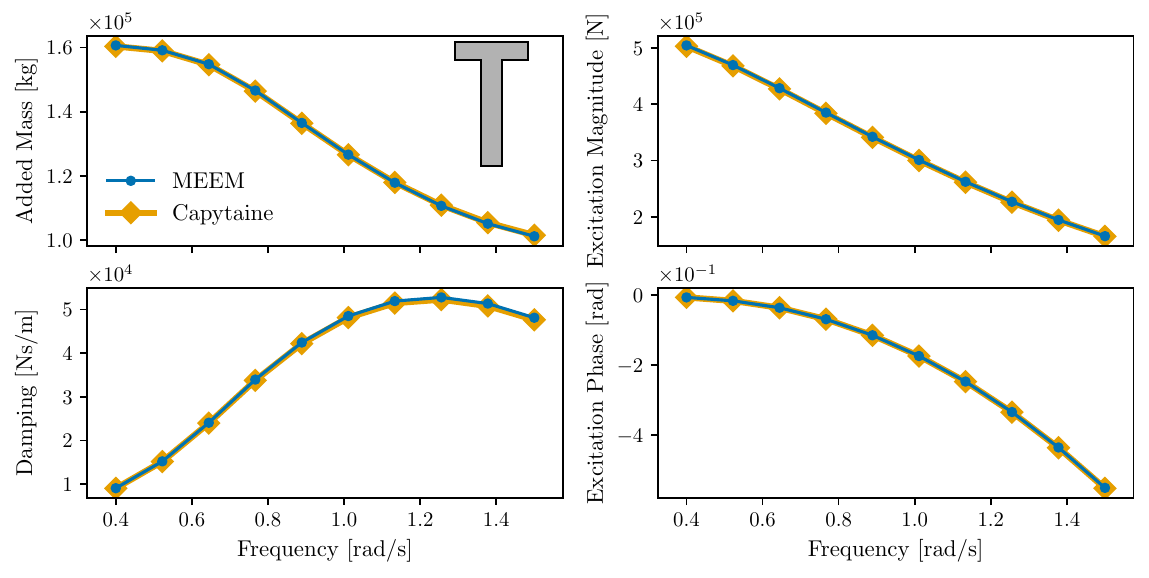}
    \caption{Added mass, radiation damping, excitation magnitude, and excitation phase from MEEM and Capytaine for CorPower-like WEC without slanted portions. Figure context: \href{https://calkit.io/symbiotic-engineering/openflash/figures?path=pubs\%2FJFM\%2Ffigs\%2FMEEM_vs_Capytaine_Nonslant_Validation.pdf}{Calkit} and \href{https://cocalc.com/rmccabe/MEEM/meem}{CoCalc}.}
    \label{fig:hydro coeff validation}
\end{figure}

The added mass $A_{11}(\omega)$, radiation damping $B_{11}(\omega)$, excitation magnitude $|X_1(\omega)|$, and excitation phase $\angle X_1(\omega)$ 
from MEEM and Capytaine, an open-source BEM software, are compared over a range of frequencies for a heaving single-body CorPower-like WEC geometry from \citet{faedo2021energy}. 
This is shown in Fig.~\ref{fig:hydro coeff validation}. 
Note that the geometry is simplified in the MEEM and Capytaine models so that it consists of two heaving surface-piercing compound annual cylinders that are rigidly fixed to one another. 
In reality, the CorPower WEC geometry also consists of a slanted section. 
This will be accounted for later in Sec.~\ref{sec:slant}. 
The geometry settings of this comparison were set as follows: 
$a_1=1.25 \text{~m}$, $a_2=4.2 \text{~m}$, $d_1=14.45 \text{~m}$, $d_2=2.05 \text{~m}$, and $h=50 \text{~m}$. 
The series in each region was truncated to 150 for MEEM ($N^{i_1}=N^{i_2}=N^{e}=150$) and 1060 panels were used in Capytaine. 
The hydrodynamic coefficients from MEEM and Capytaine are on average within 1\% of one another. 
Small discrepancies can be attributed to Capytaine needing more panels to be completely converged, a consequence of Capytaine's slow convergence with the number of panels. 
In Sec.~\ref{sec:convergence}, the dependence of MEEM's convergence on the geometry will be discussed more in depth, 
and the trade-offs between accuracy and computation time for MEEM and Capytaine will be compared in Sec.~\ref{sec:compute-time}.

\section{Convergence of Hydrodynamic Coefficients}\label{sec:convergence}
Characterizing the convergence behaviour of numerical solvers is crucial for ensuring accuracy while avoiding unnecessarily long runtimes. 
For BEM solvers, a mesh convergence study is typically performed, where the number of panels is increased until a desired level of accuracy of the hydrodynamic coefficients is achieved. 
As the matrix size of BEM is equal to the number of panels, improving accuracy through finer meshes comes with added computational cost. 
In the case of MEEM, the behaviour is very similar, except instead of increasing the number of panels, the number of eigenfunction terms $N_\mathrm{T} = N^{i_1} + 2\sum_{m=2}^M N^{i_m} + N^e$ can be increased. 
Furthermore, higher values of $N_\mathrm{T}$ imply larger matrix sizes, leading to longer runtime. 
The goal of this convergence study is to determine the number of terms $N^{i_m}$ per region needed to achieve a desired maximum error $\epsilon$ in the hydrodynamic coefficients solely based on the geometry and wave conditions being modelled. 

\begin{figure}
    \centering
    \includegraphics[width=0.9\linewidth]{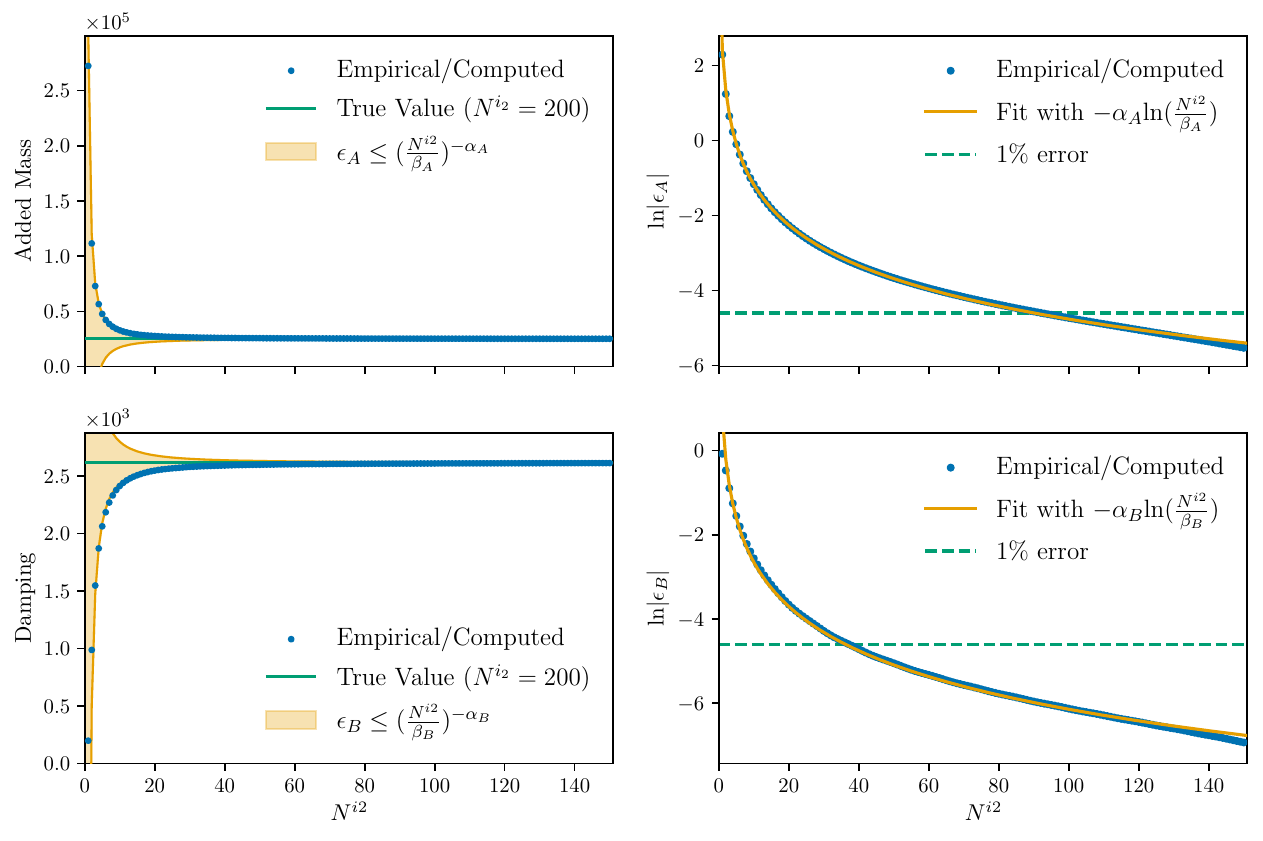}
    \caption{Left: Added mass and damping calculated for a three body region configuration at $N^{i_1} = N^{i_3} = N^e = 200$ with region $i_2$ heaving, for varying $N^{i_2}$. 
    Right: The data at left is transformed to the natural log of the associated error, 
    and fitted to obtain error envelope parameters $\alpha, \beta$ for each of added mass and damping. Figure context: \href{https://calkit.io/symbiotic-engineering/openflash/figures?path=pubs\%2FJFM\%2Ffigs\%2Falpha-beta-explanation.pdf}{Calkit} and \href{https://cocalc.com/rmccabe/MEEM/meem}{CoCalc}.}
    \label{fig:error-model}
\end{figure}

This study makes two simplifying assumptions about the behaviour of the convergence, which are found to be reasonably accurate for the configurations considered. 
First, the number of terms needed in each region depends directly on dimensionless parameters (Sec.~\ref{sec:finding-dimensionless-parameters}) derived from the geometry and not on the number of terms needed in any other region (up to correlations from sharing the same overall geometry variables). 
This is as opposed to needing to raise a region's base term count, dependent on its own geometry, to support neighboring regions with higher requirements. 
Second, we model the error $\epsilon$ for $N^{i_m}$ terms in region $m$, assuming all other regions are sufficiently converged, as a power-law envelope of $\epsilon \approx (\frac{N^{i_m}}{\beta^{i_m}})^{-\alpha^{i_m}}$ (Fig.~\ref{fig:error-model}), with the dimensionless parameters informing $\alpha^{i_m}, \beta^{i_m} > 0$. 
Dependencies differ between the two hydrodynamic coefficients, so each is modelled separately with its own final $\alpha^{i_m}_A, \beta^{i_m}_A$ (added mass) or $\alpha^{i_m}_B, \beta^{i_m}_B$ (damping). 
As the excitation force is a result of the radiation problem, due to the Haskind relation, convergence of the added mass and radiation damping is assumed to imply the convergence of the excitation force. 
Convergence is faster for larger $\alpha^{i_m}$ and smaller $\beta^{i_m}$. 

\begin{figure}
    \centering
    \includegraphics[width=0.9\linewidth]{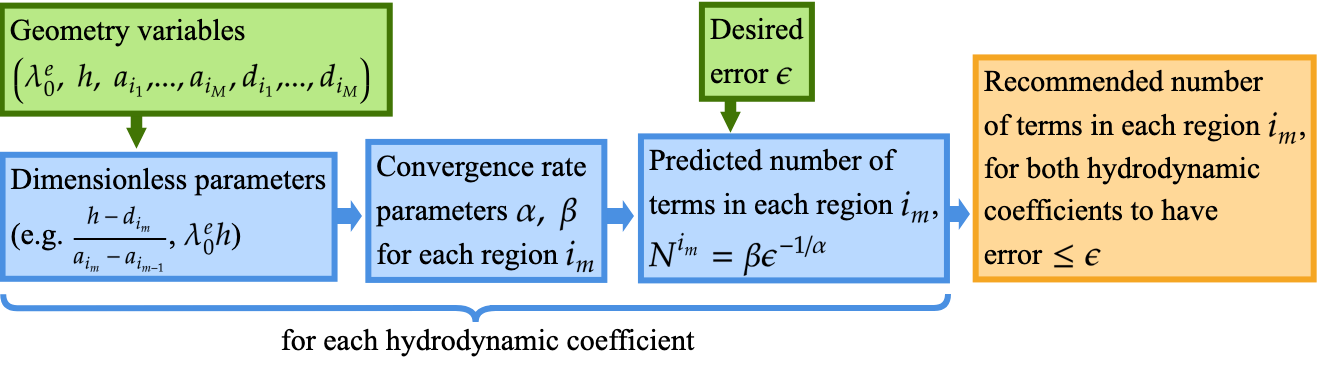}
    \caption{Flowchart of term count prediction. A geometry and desired error (green) are passed into the formula (blue) determined by the convergence study, producing a term count recommendation (orange).}
    \label{fig:nmk-predict-flowchart}
\end{figure}

After confirming these assumptions, the main objective of the convergence study is to find key dimensionless parameters for a given problem, and find $\alpha$ and $\beta$ as a function of these dimensionless parameters. 
Once this analysis is complete, the number of terms needed in each region can be predicted by taking the maximum of the number required to converge added mass and radiation damping.
This process is shown in Fig.~\ref{fig:nmk-predict-flowchart}.

\subsection{Convergence Study Procedure} \label{sec:finding-dimensionless-parameters}
The set of dimensionless parameters that characterize a region's convergence can be divided into local and cumulative parameters. 
Local parameters are extracted from the geometry of the region itself, such as its fluid height, radial width, or the ratio of its fluid height to those of its neighbors. 
Cumulative parameters are overall metrics combining the geometries of other regions, such as total radial distance from the centre or exterior region. 
Neither parameter type should depend on a specific non-neighboring region or the number of other regions, because their form should be generalizable to a configuration with any number of regions. 
Sets of dimensionless parameters are not unique, but good sets will have more direct dependencies on each parameter individually.

We categorize regions into four types by their local geometry: innermost ($i_1$), middle ($i_{2\leq m < M}$), outermost ($i_M$), and exterior ($e$). 
The local geometry of the $i$-type regions differ by region type ($r=0$, body region, or exterior region) of their inner and outer neighbors. 
To capture these region types, the majority of configurations we generate convergence data for have three body regions and one exterior region (i.e. $M = 3$).
In this study, the single cylinder case with an inner boundary of $r=0$ and outer neighbour of the exterior region is not considered.

To obtain numerical values for the convergence of a configuration's hydrodynamic coefficients with respect to $N^{i_m}$, 
they are first computed with all $N^{j}$ ($j\in\{i_1,...,i_M, e\}$) at a high number $N^{max} = 200$ for their ``true values.'' 
Then they are recomputed with all $N^{j}, j\neq i_m$ fixed to $N^{max}$, while, for the $m$th region, which we will refer to as the target region, 
$N^{i_m}$ is varied over $[1, 2, \dots, N^{big}]$, where $N^{big} = 150$ is less than $N^{max}$ by a reasonably large margin. 
From this, the errors $\epsilon = \frac{\text{value}-\text{true}}{\text{true}}$ of each hydrodynamic coefficient at each $N^{i_m}$ are calculated.

We observe that for a fixed geometry, convergence is slower 
(1) for added mass than damping in innermost and middle regions, 
(2) for body regions when the targeted region is the only one heaving (as opposed to any other combination of heave states), and 
(3) for the exterior region when the outermost region is the only one heaving. 
The remainder of this convergence study is restricted to these upper bound cases. 
In particular, claims about a region's convergence are in the context of it being the only one heaving (for body regions) or only the outermost region heaving (for the exterior region).

For each region, we begin by generating convergence data for random geometries, plotting the $N^{i_m}$ of $1\%$ convergence against theorized dimensionless parameters, and identifying the one with the clearest correlation. 
We then fix the dominant dimensionless parameter and repeat the process to find the next most important parameter. 
This is repeated until a set of dimensionless parameters capturing most of the variation is found 
(i.e. for random geometries with all of them fixed, the set of $1\%$ convergence points have sufficiently low variance).

\begin{figure}
    \centering
    \includegraphics[width=0.9\linewidth]{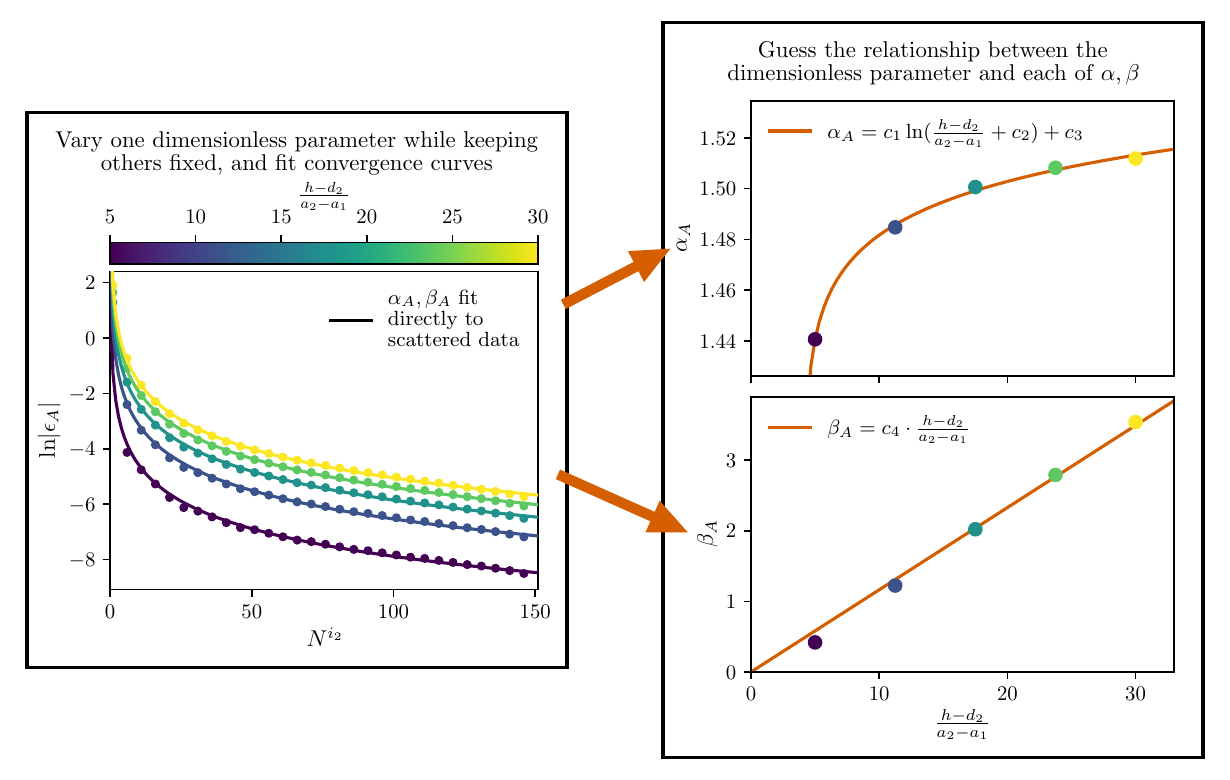}
    \caption{Process of choosing the fitting function for each dimensionless parameter (here $\frac{h-d_2}{a_2 - a_1}$), once identified. Figure context: \href{https://calkit.io/symbiotic-engineering/openflash/figures?path=pubs\%2FJFM\%2Ffigs\%2Falpha-beta-to-trend.pdf}{Calkit} and \href{https://cocalc.com/rmccabe/MEEM/meem}{CoCalc}.}
    \label{fig:choosing-fitting-functions}
\end{figure}

Next, we sweep each identified parameter over a range of reasonable values while keeping all other dimensionless parameters fixed. 
These isolated effects are quantified by plotting the $\alpha^{i_m}$ and $\beta^{i_m}$ fitted to each geometry against the dimensionless parameter. 
These plots suggest functional forms for the relationship between the dimensionless parameters and fit parameters (Fig.~\ref{fig:choosing-fitting-functions}).

Finally, functions incorporating all relevant dimensionless parameters in a region (typically a product of the functional forms of individual parameters) are theorized and assessed by fitting them to sets of randomized geometry and their associated $(N^{i_m}, \epsilon)$ pairs. 
Functions explored have to remain positive on the domain of the dimensionless parameter, restricting their form. 
For example, some relationships appear nearly linear with negative slope on the range of the dimensionless parameter they are observed at, but are expressed as decaying exponentials to avoid straying below zero.

\subsection{Final Models}
\begin{table}
\centering \begin{tabular}{|c|c|c|c|}
\hline
&$p = \frac{h-d_m}{a_m - a_{m-1}}$
&$p = \frac{h-d_{m+1}}{h-d_m}$
&$p = \frac{h-d_{m-1}}{h-d_m}$\\
\hline
$\alpha_A^{i_{m=1}}$
& $c_1\ln(p + c_2) + c_3$
& $\begin{cases}
    c_1, & p < 1\\
    c_2, & p \geq 1
\end{cases}$ & N/A\\
\hline
$\beta_A^{i_{m=1}}$
& $c_1\cdot p$
& $\begin{cases}
    \frac{c_1}{1 + e^{-c_2(p-1)}} + c_3, & p < 1\\
    c_4, & p \geq 1
\end{cases}$
& N/A\\
\hline
$\alpha_A^{i_{2 \leq m < M}}$
& $c_1\ln(p + c_2) + c_3$
& $\begin{cases}
    c_1, &p < 1\\
    c_2, &p \geq 1
\end{cases}$
& $\begin{cases}
    c_1, &p < 1\\
    c_2, &p \geq 1
\end{cases}$ \\
\hline
$\beta_A^{i_{2 \leq m < M}}$\
& $c_1\cdot p$
& $\begin{cases}
    \frac{c_1}{1 + e^{-c_2(p-1)}} + c_3, & p < 1\\
    c_4, & p \geq 1
\end{cases}$
& $\begin{cases}
    c_1, &p < 1\\
    c_2, &p \geq 1
\end{cases}$\\
\hline
\end{tabular}
\begin{tabular}{|c|c|c|c|c|c|}\hline
&$p = \frac{h-d_M}{a_M - a_{M-1}}$
&$p = \frac{h}{h-d_M}$
&$p = \frac{h-d_{M-1}}{h-d_M}$
&$p = \frac{a_M - a_{M-1}}{a_{M-1}}$
&$p = \lambda_0^e h$\\
\hline
$\alpha_A^{i_M}$
& $c_1\ln(p + c_2) + c_3$
& $c_1\ln(p + c_2 - 1) + c_3$
& $\begin{cases}
    c_1, & p < 1\\
    c_2, & p \geq 1
\end{cases}$
& Independent
& Independent\\
\hline
$\beta_A^{i_M}$
& $c_1\cdot p$
& $c_1\ln(p + c_2 - 1) + c_3$
& $\begin{cases}
    c_1, & p < 1\\
    c_2, & p \geq 1
\end{cases}$
& $c_1\cdot p + c_2$
& Independent\\
\hline
$\alpha_B^{i_M}$
& Independent
& $c_1\ln(p + c_2 - 1) + c_3$
& Independent
& Independent
& $c_1e^{-c_2p} + c_3$\\
\hline
$\beta_B^{i_M}$\
& $c_1\cdot p + c_2$
& $c_1e^{-c_2p}$
& Independent
& Independent
& $\begin{cases}
    c_1 \frac{p}{\sqrt 5}, & p < 5\\
    c_1 \sqrt p, & p \geq 5
\end{cases}$\\
\hline
\end{tabular}
    \caption{Dependency trends on individual dimensionless parameter identified/used for fitting error predictions in body regions, based on the procedure in Fig.~\ref{fig:choosing-fitting-functions}. 
    The constants $c_i \geq 0$ in each box are placeholders and not intended have the same value between boxes. 
    Damping nearly always converges faster than added mass in the interior regions ($m<M$), so the associated fit parameters ($\alpha_B^{i_m}$, $\beta_B^{i_m}$) were not analysed. 
    The choice of the piecewise form of $\beta_B^{i_M}(\lambda_0^eh)$ is out of practicality for fitting: 
    to capture the clear $\beta_B^{i_M} \propto \sqrt{\lambda_e^0 h}$ relation at high $\lambda_e^0 h$ while approximating the different behaviour near zero.
    }
    \label{tab:final-fitting-functions}\end{table}
Approximate dependencies between fit parameters and dimensionless parameters in the body regions are listed in Table~\ref{tab:final-fitting-functions}. In this section, we interpret some of the most influential dimensionless parameters.

The number of terms in a region's eigenfunction expansion necessary to sufficiently resolve the velocity potential's detail there (i.e. achieve convergence) depends on the detail itself and the resolving power. In other words, influential dimensionless parameters relate the spatial frequency of the detail (variation in the true value of the potential) to the characteristic length of a region's eigenfunctions.

In the body region $i_m$, the most important parameter is its $\frac{\text{fluid height}}{\text{radial width}} = \frac{h - d_{m}}{a_{m} - a_{{m-1}}}$. The radial eigenfunctions here inherit their eigenvalues $\lambda_n^{i_m} = \frac{n\pi}{h-d_m}$ from the corresponding vertical eigenfunction. Thus, taller regions need higher harmonics to create the same characteristic length of resolution for radial details, and the ratio of fluid height to radial width is a ratio of resolving scale to detail.

In the exterior region, the analogous relationship between the two major characteristic lengths is captured by $\lambda_0^eh$. Convergence slows as $\lambda_0^eh$ increases, particularly in the case of damping. This trend is consistent with Sec.~\ref{sec:low-freq}, which shows that only one term per region (exterior or otherwise) is needed when $\lambda_0^e$ is near zero.

In the body regions, another notable local parameter is the ratio of region's neighbour's fluid heights to that of its own. We observed that convergence is significantly faster for region $i_m$ when $\frac{h-d_{m-1}}{h-d_m} < 1$ than when it's greater than $1$. The convergence speed changes quickly (almost step-like) near $1$. The trend is the same for $\frac{h-d_{m+1}}{h-d_m}$.

This is theorized to result from the boundary condition represented by the ratio. If $\frac{h-d_{m-1}}{h-d_m} < 1$, then region $m$'s entire inner boundary condition is the continuity conditions between regions $m$ and $m-1$, but if $\frac{h-d_{m-1}}{h-d_m} > 1$, the boundary includes a section of $\frac{\partial \phi}{\partial r} = 0$ near $z = d_m$, where the hydrodynamic coefficients are being integrated. This constant rather than matching boundary condition may create simpler spatial details, leading to faster convergence.

A cumulative parameter observed for the innermost region ($i_1)$ was some dependence on the depth of the outermost region ($i_3$), where convergence for $N^{i_1}$ was faster if $d_3 > d_2$, which we termed a shielding effect. 
We removed this complicating factor by only considering the upper bounding case $d_3 < d_2$ when predicting for the innermost region (e.g. the data set for $i_1$ in Fig.~\ref{fig:convergence-fit-assessment}). 
However, this suggests the possibility that the impact of fluid height ratios, especially in the case of the outer neighbour, might be a special case of the shielding effect.

In the exterior region, other influential parameters include $\frac{h-d_M}{a_M-a_{M-1}}$ and $\frac{h-d_M}{h}$ for both hydrodynamic coefficients and $\frac{a_M}{h}$ for added mass only. Convergence was slower for larger $\frac{h-d_M}{a_M-a_{M-1}}$ (the outermost region's main resolution-detail ratio) and smaller $\frac{a_M}{h}$. However, the convergence relation with $\frac{h-d_M}{h}$ was complicated by large oscillatory variation in the hydrodynamic coefficients with respect to increasing $N^{i_m}$ (i.e. the configurations exhibited significant nonmonotonic convergence) for values of $\frac{h-d_M}{h}$ near $1$. By $\frac{h-d_M}{h} < 0.8$ though, there is a clear trend of slower convergence for smaller $\frac{h-d_M}{h}$.

\subsection{Assessment} \label{sec:convergence-fit-assessment}

\begin{figure}
    \centering
    \includegraphics[width=0.9\linewidth]{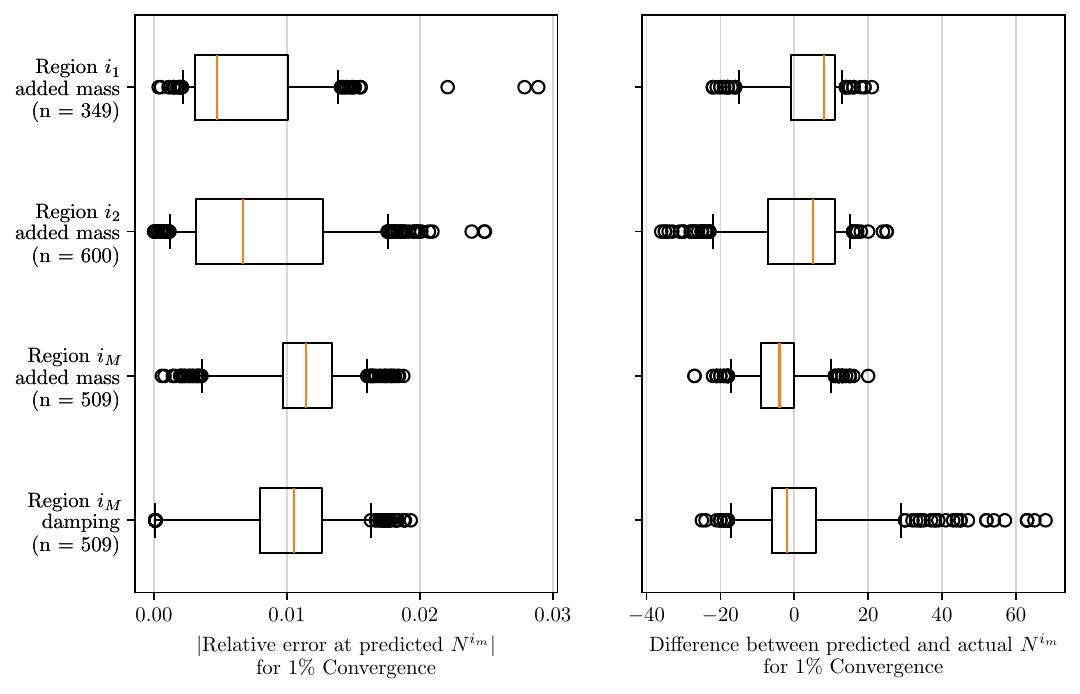}
    \caption{Randomly generated (described in Sec.~\ref{sec:convergence-fit-assessment}) configurations with the target region heaving were fit with the product of the dependencies in Table~\ref{tab:final-fitting-functions}; 
    the accuracy of the resulting fits for predicting $1\%$ error are shown. 
    Note that the whiskers are at the 5th and 95th percentile, $n$ here is the size of each set of configurations, and configurations were selected that converged by $N^{i_m} = 150$. Figure context: \href{https://calkit.io/symbiotic-engineering/openflash/figures?path=pubs\%2FJFM\%2Ffigs\%2Fconvergence-fit-assessment.pdf}{Calkit} and \href{https://cocalc.com/rmccabe/MEEM/meem}{CoCalc}.} 
    \label{fig:convergence-fit-assessment}
\end{figure}

The final fits chosen are assessed in Fig.~\ref{fig:convergence-fit-assessment}. 
The $\alpha^{i_m}$ and $\beta^{i_m}$ calculated from the dimensionless parameters are used to predict the $N^{i_m}$ of $1\%$ convergence, and that $N^{i_m}$ and associated error compared to reality. 
The configurations in those sets were generated from uniformly at random choosing: 
$1<\lambda_0^eh<80$, three $1<\frac{h-d_m}{a_m-a_{m-1}}<30$, and three $0.05 < \frac{d_m}{h} < 0.95$ for a three body configuration. 
Also, all fits had the additional condition that if the prediction was less than $N^{i_m} = 20$, it was changed to $20$ to avoid egregiously high errors if the prediction was wrong. 
In reality, many configurations can achieve $1\%$ convergence before $N^{i_m} = 20$.

As shown in Fig.~\ref{fig:convergence-fit-assessment}, the fits provide acceptable amounts of error and predictions of the required $N^{i_m}$. 
In over $95\%$ of cases, hydrodynamic coefficients are modelled with less than $2\%$ error and the overprediction of $N^{i_m}$ does not exceed $30$ terms. 
In practice, this means if the models produced by fitting the data sets in Fig.~\ref{fig:convergence-fit-assessment} were used to predict the terms needed to model some new configuration within $1\%$ accuracy, 
the actual accuracy would almost certainly be better than $3\%$, and better than $1.5\%$ over three quarters of the time. 
Meanwhile, $N^{i_m}$ overestimates are in the tens even though the necessary amount could be over $100$, saving computation time over setting $N^{i_m}$ to $200$ in all regions. 
More conservative predictions can be obtained by adding weights in the fitting stage to penalize underestimates of $N^{i_m}$, or modifying $\alpha^{i_m}$, $\beta^{i_m}$, and/or the final number of terms by some constant value.

The main limitations of this convergence study stem from the simplifying assumptions (Sec.~\ref{sec:convergence}). 
Beyond these, another complication encountered is that some convergences exhibit decaying oscillations about the true value (as opposed to the monotonic behaviour in Fig.~\ref{fig:error-model}), 
obfuscating the envelope's relative errors and $1\%$ convergence point, and ability to measure the true fit parameters. 
Furthermore, in the outermost region we observe that the convergence initially follows one logarithmic curve, but then moves along another, steeper one at a particular $N^{i_M}$, rendering the first curve an overestimate. 
Our formulation tends to fit to the first curve, which is why the associated box plot in Fig.~\ref{fig:convergence-fit-assessment} has more significant overestimates than the other regions do. 

Overall, the results of the convergence study enable users to accurately compute hydrodynamic coefficients without added computation time. 
Determining sufficient parameter settings before modelling is particularly useful in the field of optimization, where the hydrodynamic coefficients for a wide variety of geometries may be computed numerous times. 
By characterizing the convergence behaviour a priori and using the predicted number of terms, users can perform optimization with confidence that their model's accuracy is acceptable.

\section{Slanted Geometries}\label{sec:slant}
So far, we have only considered geometries in the form of a series of surface-piercing compound annular cylinders, as shown in Fig.~\ref{fig:Diagram}. 
In this section, we will discuss how MEEM can be used to approximate the first-order hydrodynamic forces on vertically axisymmetric surface-piercing bodies with radially-monotonic profiles. 
This generalization is important for modelling body shapes that are seen in engineering applications. 
The CorPower WEC \citep{de2016techno}, WaveBot WEC \citep{strofer2023control}, AquaHarmonics WEC \citep{weaver2020super}, LUPA WEC float \citep{beringer2025degrees}, and RM3 WEC float \citep{neary2014methodology} are all examples of geometries that can be modelled with this extension. 
In literature, toroidal \citep{mavrakos1997hydrodynamic} and spherical bodies \citep{zhang_hydrodynamic_2016} have also been modelled using MEEM. 
However, little discussion has been given on the accuracy of this approximation, how slanted geometries should be discretized, and how the steepness of the slanted geometry influences the convergence of the MEEM solution. 
In this section, we will address these topics.

\subsection{Discretization}

\begin{figure}[htbp]
    \centering
    \includegraphics[width=0.95\linewidth]{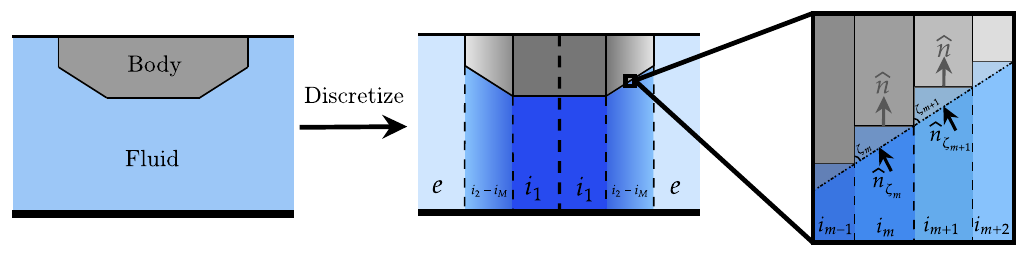}
    \caption{Cross-sectional view of a slanted geometry and its equivalent discretized geometry. The close-up shown on the right-hand side shows how a slanted region can be approximated by a finite number of cylindrical rings. The unit vector normal to the horizontal part of all cylindrical rings is $\hat{n}$, while the unit vector normal to the true slanted body in the $m$th region is $\hat{n}_{\zeta_m}$.}
    \label{fig:Discretization Diagram}
\end{figure}

\begin{figure}[htbp]
    \centering
    \includegraphics[width=0.95\linewidth]{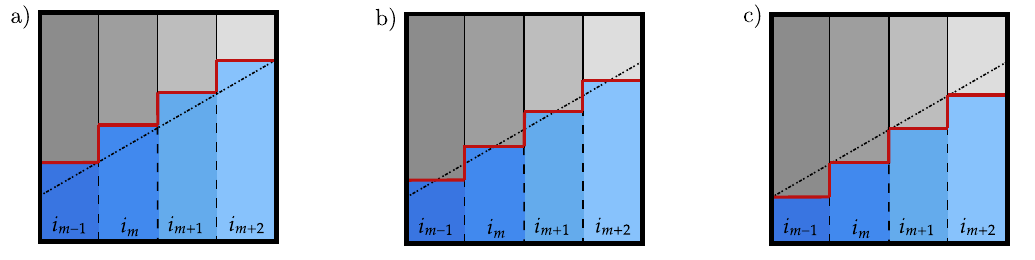}
    \caption{Cross-sectional view of different discretization schemes for approximating a slanted geometry with cylindrical rings. The geometry of the cylindrical rings can be chosen such that their horizontal surfaces are a) within, b) partially within, or c) outside the true slanted body shape.}
    \label{fig:Discretization Schemes Diagram}
\end{figure}

As formulated, MEEM can only exactly represent geometries where the fluid can be divided into regions that have the top boundary parallel to the sea floor. The standard way to use MEEM with slanted or curved bodies is to approximate the geometry by subdividing slanted regions into numerous stepped regions approximating the true outline~\citep{kokkinowrachos_behaviour_1986}, as shown in Figure~\ref{fig:Discretization Diagram}. There are many options for this discretization. As shown in Fig.~\ref{fig:Discretization Schemes Diagram}, the wetted surface of the discretized body modelled can lie a) entirely within, b) partially within, or c) outside of the slanted wetted surface that is being approximated. In cases a) and c), the computed potentials in the true fluid region will be smooth. Alternatively, in case b), the approximating outline can cross back and forth across the true outline, meaning some points in the true fluid are inside of the body. This creates discontinuities in the potential at points in the true fluid region that correspond to a fluid-body boundary in the approximation. Thus, in this work, the potential is not integrated over the true slanted wetted surface. Instead, the pressure is integrated over the discretized wetted surface, outlined in red in Fig.~\ref{fig:Discretization Schemes Diagram}.

However, this discretization quickly becomes computationally expensive. Not only does matrix size increase as the number of regions increases, but the number of terms required for convergence in each subregion also increases. This is because increasing the number of subdivisions reduces the radial width allocated to each subregion without significantly changing the fluid height, increasing $\frac{\text{fluid height}}{\text{radial width}}$ (see Section~\ref{sec:convergence}).

A complicating factor is that regions with shallower slants can be represented more accurately with fewer subdivisions than those with steeper slants. Another is that the hydrodynamic forces on the entire body can be thought of as a weighted sum of the forces on each region, where the radial width of each region determines the weights. 
Since the forces applied to regions with smaller radial widths get weighted less, MEEM is able to accurately determine the hydrodynamic coefficients of bodies with poorly approximated slanted regions when the slanted regions have small radial widths.

\subsection{Sources of Inaccuracies}
When modelling a slanted geometry as a discretized one using MEEM, inaccuracies can accrue when 1) computing the radiated velocity potential and 2) integrating pressure over the discretized wetted surface. Since the radial and vertical dependence was separated in Table~\ref{tab:MEEM-eigenfunctions}, coupling between the radial and vertical directions cannot be captured. Additionally, the boundary condition on the slanted wetted surface will not be satisfied when using MEEM on the discretized geometry. This leads to discrepancies in the velocity potential. Furthermore, integrating the pressure over the discretized wetted surface instead of the slanted wetted surface leads to additional inaccuracies since the pressure is not evaluated at the correct spatial points. If one considers integrating the pressure on a slanted surface $S_p$ at constant angle $\zeta$, instead of over the horizontal sections of the discretized geometry, $\hat n_p \cdot \hat e_z = \sin(\zeta)$ and $dS = \frac{r}{\sin(\zeta)}drd\theta$ in Eq.~\ref{eq:freq domain scalar rad force}. When these are substituted, the $\sin(\zeta)$ cancels and the result is the same as Eq.~\ref{eq:freq domain scalar rad force in terms of regions}, except $d_m$ now has $r$ dependence. However, the implicit dependence of the $Z$ eigenfunction evaluations on $r$ means that the integrals including homogeneous potentials cannot be simplified as in Eqs.~\ref{eq:freq domain scalar rad force in terms of added mass and damping} and onward. Unless the region's outline is flat and $d_m(r)$ is constant, the hydrodynamic coefficient integrals (given $\vec x$) do not admit closed forms. 

Instead of altering the MEEM procedure for slanted geometries, the same procedure can be used for modelling slanted geometries with the assumption that, as the number of subdivisions increases, the discretized geometry approaches the true geometry, and, consequently, inaccuracies in the hydrodynamic coefficients diminish. Thus, a proper choice of the number of subdivisions is required to mitigate these inaccuracies.

\subsection{Inaccuracies in Velocity Potential} 
\begin{figure}
    \centering
    \includegraphics[width=0.9\linewidth]{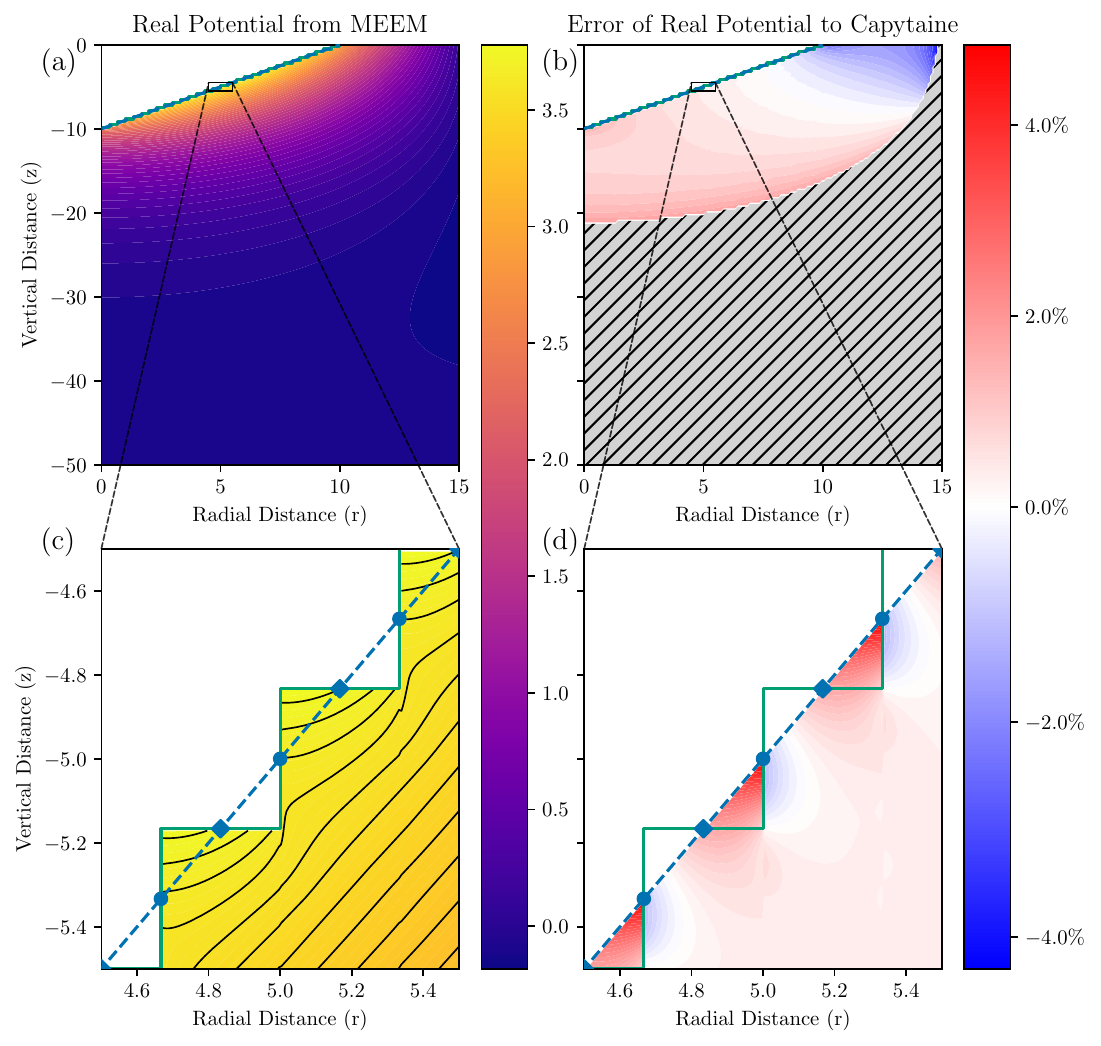}
    \includegraphics[width=0.8\linewidth]{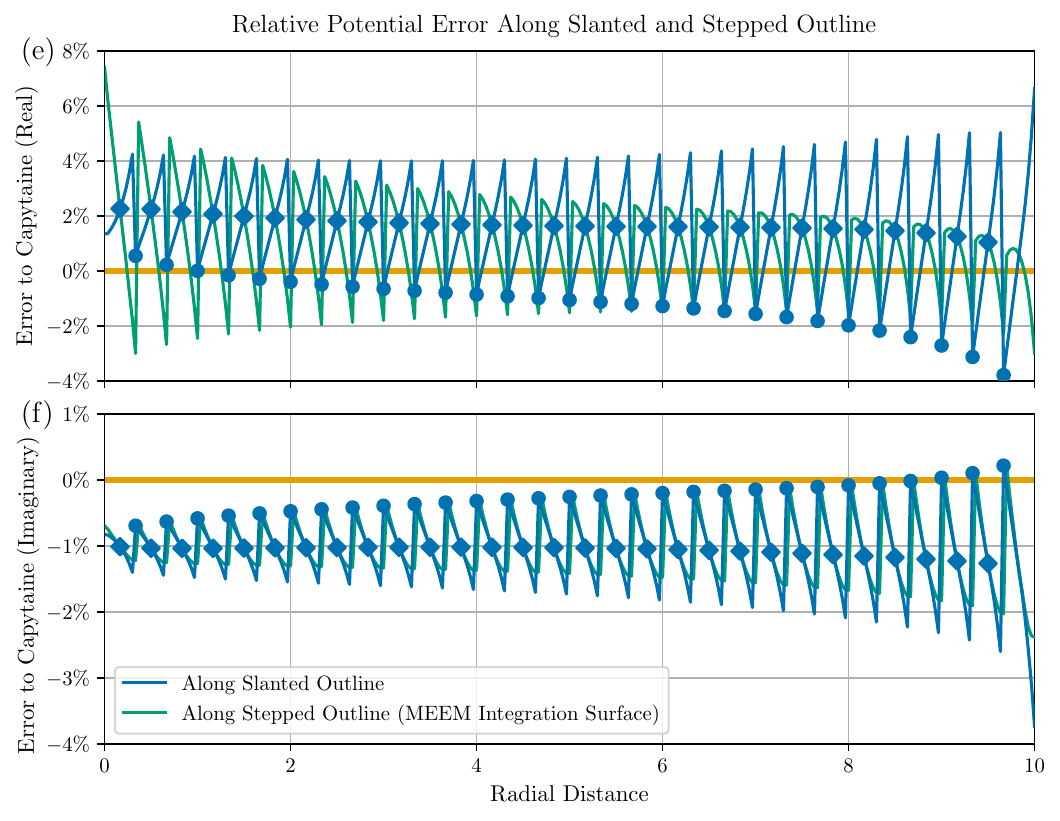}
    \caption{Comparison of the radiated velocity potential computed from MEEM and Capytaine. (a) and (c) show the real part of the potential from MEEM, (b) and (d) show the error in the real potential from MEEM relative to Capytaine, and (e) and (f) show the error in the potential along the slanted and stepped outlines. Figure context: \href{https://calkit.io/symbiotic-engineering/openflash/notebooks?path=analysis\%2Fslant_data_plots.ipynb}{Calkit} and \href{https://cocalc.com/rmccabe/MEEM/meem}{CoCalc}.} 
    \label{fig:contour potential comparison plots}
\end{figure}

While the hydrodynamic coefficients computed using MEEM may approach their true values for a slanted geometry, inaccuracies in the velocity potential are still expected locally. Fig~\ref{fig:contour potential comparison plots} (a) and (b) show the real value and error in the real value of the radiated velocity potential for a discretized body (modelled with MEEM) relative to a slanted body (modelled with Capytaine). Note the error in the hatched region in Fig~\ref{fig:contour potential comparison plots} (b) is removed as the value of the velocity potential is small in this region ($\mathrm{Re}(\phi)<0.2$). There is an overall trend of error moving from positive to negative as the depth becomes shallower. Fig~\ref{fig:contour potential comparison plots} (c) and (d) show a more detailed view of the local behaviour of the velocity potential. The contours of constant velocity potential, indicated as solid black lines in Fig~\ref{fig:contour potential comparison plots} (c), either start and end at vertical and horizontal sections of the discretized surface, or are parallel to the slant that is being approximated, indicated as dashed blue lines in Fig~\ref{fig:contour potential comparison plots} (c). Fig~\ref{fig:contour potential comparison plots} (d) shows the corresponding error in the real value of the velocity potential (relative to Capytaine). The largest positive error is within the discretized body. For each step, there is positive error on the horizontal surface and negative error on the vertical surface. 

If one were to consider a different integration scheme to minimize the error in the velocity potential that is used in the hydrodynamic coefficient calculations, one may seek to sample points in Fig~\ref{fig:contour potential comparison plots} (d) where error is zero, and numerically integrate over the true slant outline. However, the precise radial and vertical coordinates where the error is zero are unknown before performing the computation and comparing with a BEM solver. A different option is to systematically sample the velocity potential at spatial points that are determined by the slant and discretized geometries. For example, one could sample the velocity potential at locations where the slant surface intersects with the discretized surface. These points are indicated in Fig~\ref{fig:contour potential comparison plots} (c-e) with dots and diamonds for points sampled along the vertical and horizontal surfaces, respectively. Fig~\ref{fig:contour potential comparison plots} (e) shows the error in the real and imaginary parts of the velocity potential along the slant and stepped (discretized) surfaces across the entire radial width. When comparing the real parts, sampling at the horizontal surfaces leads to errors less than $3\%$, while sampling at the vertical surfaces can result in close to $4\%$ error. However, when considering the imaginary parts, sampling along the vertical surface results in less error. Still, errors will be present regardless of which of these sampling locations is chosen. While accurate hydrodynamic coefficients were obtained using the MEEM formulation discussed in Sec.~\ref{sec:mathematical-formulation}, Fig~\ref{fig:contour potential comparison plots} (c-e) show that systematic approaches for sampling the potential may be effective ways to mitigate inaccuracies and reduce the number of discretizations needed.

\subsubsection{Dependence on Slant Steepness} \label{subsection: slant intro hydros}
\begin{figure}
    \centering
    \includegraphics[width=1\linewidth]{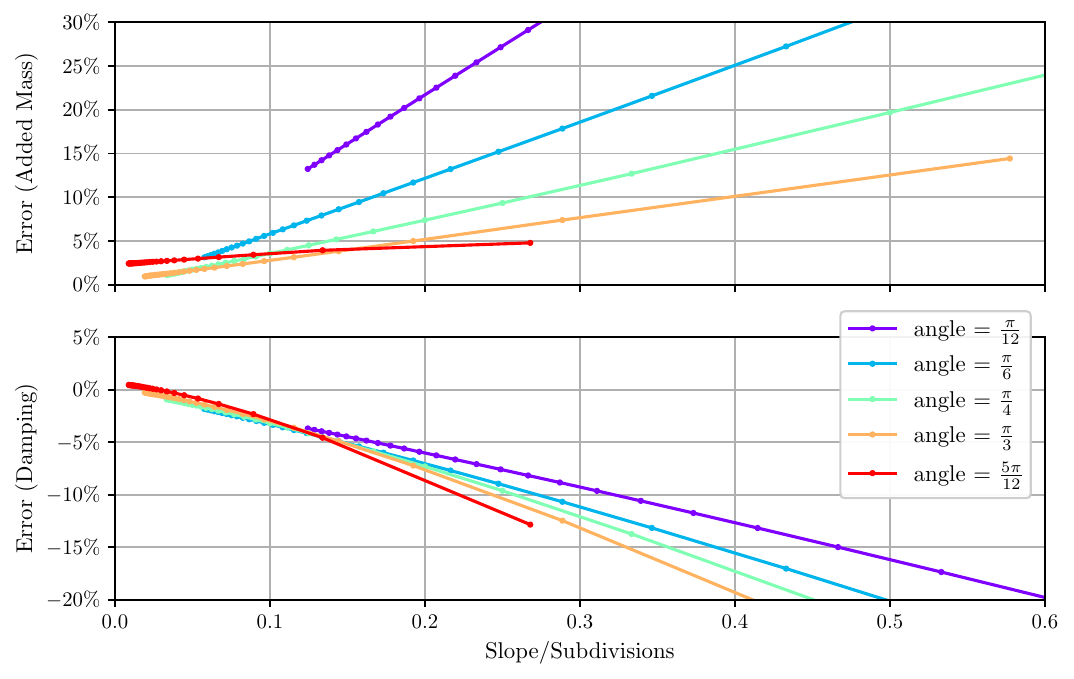}
    \caption{Convergence of the hydrodynamic coefficients of five cones of varying steepness calculated with MEEM's step approximation (radius $10$, $h=50$, $\omega = 1$, $N^{i_m} = N^e = 400$). Subdivisions calculated ranged from $1$ to $30$. Error was calculated relative to values given by Capytaine. 
    Figure context: \href{https://calkit.io/symbiotic-engineering/openflash/figures?path=pubs\%2FJFM\%2Ffigs\%2FVary-Slopes.pdf}{Calkit} and \href{https://cocalc.com/rmccabe/MEEM/meem}{CoCalc}. 
    }
    \label{fig:slope-dependence}
\end{figure}

To determine the number of subdivisions needed to accurately model a slant with a given steepness, we compare the hydrodynamic coefficients from Capytaine and MEEM for a variety of steepnesses and number of subdivisions when modelling a cone. 
As shown in Fig.~\ref{fig:slope-dependence}, steeper slants (smaller angle $\zeta$) require more subdivisions for the same level of accuracy, 
and the error in the approximation is roughly inversely proportional to the number of subdivisions. 
Convergence of the added mass with respect to the number of subdivisions is slower than that of damping, 
as shown by the added mass for an angle of $\pi/12$ (15\textsuperscript{o}) from vertical being unable to achieve less than $10 \%$ error with the number of subdivisions considered. 
Meanwhile, less than $5 \%$ error in damping is achieved for all angles when 30 subdivisions are used. 
While it is possible to achieve more accurate solutions using more subdivisions, 
the computation cost of modelling these steeper geometries will be higher due to the size of the $\mathbf{A}$ matrix growing with the number of subdivisions, 
potentially limiting MEEM's computational advantage over the more general BEM solvers.

\subsubsection{Validation} \label{subsection: slant validation}
\begin{figure}[htbp]
    \centering
    \includegraphics[width=0.95\linewidth]{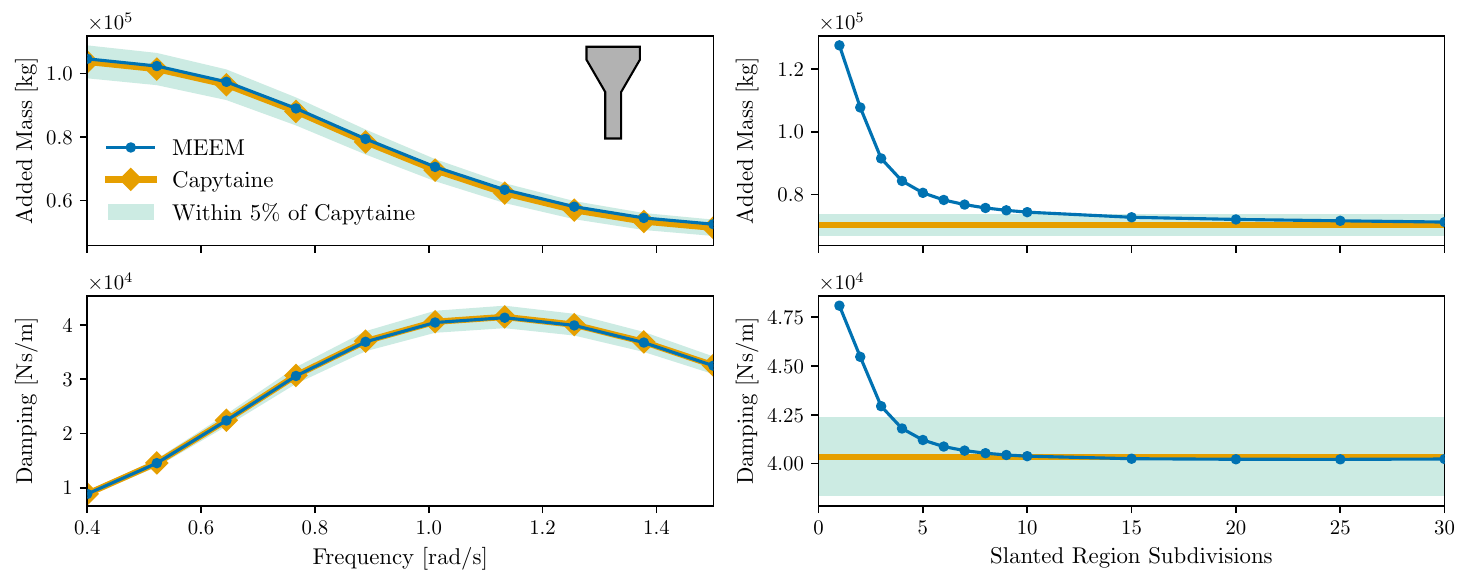}
    \caption{Left: Computed hydrodynamic coefficients for the CorPower-like WEC geometry with a slanted region (the slant intersects the vertical at $d=7.13$). MEEM approximates the slanted region using 30 subdivisions. Right: Geometry from left plots at $\omega = 1$. In both cases, MEEM uses $N^{i_m} = N^e = 400$ and Capytaine has $5940$ panels. Figure context: \href{https://calkit.io/symbiotic-engineering/openflash/figures?path=pubs\%2FJFM\%2Ffigs\%2FMEEM-CPT-Slant-Freq.pdf}{Calkit} and \href{https://cocalc.com/rmccabe/MEEM/meem}{CoCalc}.}
    \label{fig:MEEM slant comparison}
\end{figure}

The CorePower-like geometry was modelled again using MEEM, where the slant angle was accounted for by subdividing the slanted region. As shown in the left subplots of Fig.~\ref{fig:MEEM slant comparison}, when 30 subdivisions are used in the slanted portion, the hydrodynamic coefficients computed using MEEM are within 5\% of those from Capytaine for wave frequencies between 0.4 and 1.5 rad/s. For the wave frequency of 1 rad/s, the right subplots of Fig.~\ref{fig:MEEM slant comparison} show that both the added mass and radiation damping from MEEM converge to that from Capytaine as the number of subdivisions in the slanted region is increased, where the damping converges at a faster rate. These results indicate that MEEM is an alternative method for accurately modelling surface-piercing axisymmetric bodies. Furthermore, if the hydrodynamic coefficients converge for a small enough truncation order, MEEM can be a less computationally expensive alternative. In Sec.~\ref{sec:compute-time} we explore the computation costs of MEEM, and compare them to the BEM solver Capytaine.

\section{Computation Time and Accuracy}\label{sec:compute-time}
\subsection{Time Complexity}
The runtime of the MEEM method is the time required to find the eigencoefficients, then obtain the hydrodynamic coefficients from eigencoefficients. We will consider a single radiation problem, meaning a single DOF is active, and all computation complexities discussed will be in terms of the number of floating-point operations.
First, a nonlinear root-finding algorithm runs $N^e$ times to generate the $\lambda_n^e$ inputs used in the $\mathbf{A}$ matrix and $\vec{b}^q$ vector.
Then $\mathcal{O}(N_T)$ Bessel functions must be evaluated for the radial $\mathbf{A}$ matrix terms 
and $\mathcal{O}(N_T^2)$ elementary functions for the coupling integral $\mathbf{A}$ matrix terms.
The linear solve roughly scales cubically with matrix size, $\mathcal{O}(N_T^3)$.
The radial integrals for the $\vec{c}_p$ vector do not require evaluating Bessel functions with any arguments that were not already evaluated for the $\mathbf{A}$ matrix, so these evaluations are reused. Table~\ref{tab:reusable-dependence} shows the dependence of each quantity on the frequency $\omega$ and $\eta_m$, where $\eta_m=1$ if the $m$th cylindrical ring is heaving and $\eta_m=0$ if it is fixed. 

\begin{figure}[htbp]
    \centering
    \includegraphics[width=0.95\linewidth]{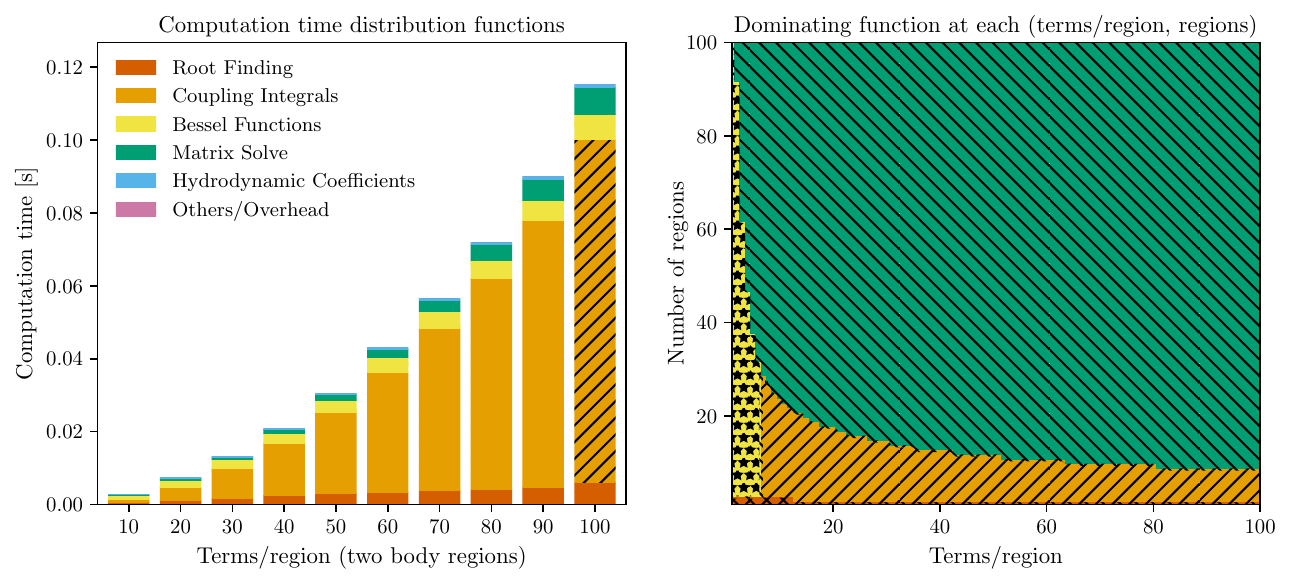}
    \caption{Left: Distribution of function computation times for the geometry in Fig.~\ref{fig:hydro coeff validation}, varying the terms per region. Right: The function taking the most computation time for combinations of region count and terms per region, assuming the same number of terms per region. Figure context: \href{https://calkit.io/symbiotic-engineering/openflash/figures?path=pubs\%2FJFM\%2Ffigs\%2FMEEM-Comp-Distribution.pdf}{Calkit} and \href{https://cocalc.com/rmccabe/MEEM/meem}{CoCalc}.}
    \label{fig:time-dominance}
\end{figure}

These time complexities are listed using more explicit per-region term counts in Table~\ref{tab:time-complexities}, and examples of their distribution/dominance given the same $N^{i_m}$ per region are given in Fig~\ref{fig:time-dominance}. For low region counts, coupling integrals dominate. For large region counts and terms per region, as is commonly encountered in slant approximations, the matrix solve will eventually dominate.

For comparison, Capytaine solves a matrix system whose side length scales with panel count, and its total runtime (for matrix element generation and matrix solve) empirically scales quadratically with panel count.

\subsection{Caching}
Computations can be cached depending on the variable changing between runs (Table~\ref{tab:reusable-dependence}). First, $\mathbf{A}$'s form is independent of whether each region is heaving or fixed ($\eta_m$), enabling the form in Eq.~\ref{A_pq B_pq matrix form} where a single $\mathbf{A}$ works with all possible $\vec b^q$ and $\vec c_p$ for the geometry (collected into the matrices $\mathbf{B}$ and $\mathbf{C}$). This implies that many of the operations that are used to solve a particular $\mathbf{A}\vec x^q = \vec b^q$ can be cached and reused as the $\eta_m$ vary.

With respect to varying $\omega$, only entries in $\mathbf{A}$ and $\vec b^q$ related to the $i_m$-$e$ region boundary are affected (and specifically, only the vertical eigenfunction components of those entries), so relatively few entries are changed when sweeping frequencies. Additionally, a common use case for hydrodynamics solvers is evaluating multiple geometries in the same environment (at the same depth and sweeping the same range of frequencies). The $\lambda_n^e$, which depend only on $h$ and $\omega$, can be reused between such runs.

\begin{table}
\centering \begin{tabular}{|l|c|c|}
\hline
& Depends on $\omega$ & Independent of $\omega$\\
\hline
Depends on all $\eta_m$ & $\vec b^q, \vec x^q$ & $\vec c_p$\\
\hline
Independent of all $\eta_m$ & $\lambda_n^e, \boldsymbol{\mathcal{Z}}^{i_M e},\mathbf{A}$& -- \\
\hline
\end{tabular}
\caption{Dependencies of various objects on $\omega$ and which regions are heaving. Changes to $\mathbf{A}$ and $\vec b^q$ due to $\omega$ only occur in rows corresponding to the boundary between regions $i_M$ and $e$.} \label{tab:reusable-dependence} \end{table}

\subsection{Future Speed Enhancements}\label{sec:speedups}
\paragraph{Matrix Solve}
Given that MEEM's matrix is sparse and its construction can be decomposed into element-wise products, there exist additional ways to speed up its solution.
In particular, the $\mathbf{A}$ matrix is block bi-diagonal, which can be solved in $\mathcal{O}(N_T)$ complexity via the Thomas algorithm, compared to the $\mathcal{O}(N_T^3)$ complexity of typical matrix solves \citep[Section 3.8.3]{quarteroni_numerical_2006}. 
However, \texttt{numpy/scipy} linear algebra solvers do not implement the Thomas algorithm in block form, so a standard direct solver is used in the present implementation.
For large matrix sizes where the linear solve dominates computation, the use of a sparse iterative solver or the Thomas algorithm could unlock computational savings, enabling the modelling of more complex geometries during expensive optimization studies.

\citet{chau2012inertia} also discuss an optional convention difference, first suggested by \citet{mavrakos_hydrodynamic_2004}, to renormalize the radial eigenfunctions in each region so they evaluate to either zero or one at boundaries.
This increases the sparsity of the $\mathbf{A}$ matrix and reduces the number of coupling integral calculations and Bessel function calculations required to obtain the eigencoefficients.
However, it does not reduce the number of Bessel calculations required to obtain the hydrodynamic coefficients because the avoided $\mathbf{A}$ matrix Bessel evaluations must still be performed in the $\vec c$ vector.
Future work could examine whether other renormalizations exist that avoid calculations overall. 

As appendix~\ref{sec:Forms of Matrix A} shows, the $\mathbf{A}$ matrix is highly structured and can be expressed as an element-wise product of a radial and a vertical factor.
Further application of linear algebra concepts could potentially yield alternative matrix expressions that would reduce computation time and enhance theoretical insights.
For example, the Hadamard-Kronecker mixed product property, Kronecker inverse, and blockwise matrix inversion could be explored to potentially obtain an analytical expression for $\mathbf{A}^{-1}$, while low-rank approximation theory could perhaps augment the experimental results of the convergence study with theoretical guarantees.

If hydrodynamic coefficients must be computed over a range of frequencies with some freedom over the exact frequency values used, one way to reduce the number of Bessel evaluations is to select the frequency vector such that some Bessel-K arguments are identical between the interior and exterior radial eigenfunctions $R_{2n_m}^{i_m}$ and $R_{1n_e}^{e}$. This occurs when
\begin{equation}\label{eq:equal-args}
\lambda_{n_m}^{i_m}a_m = \lambda_{n_e}^{e}a_M
\end{equation}
for any $(n_m,n_e,m)$ positive integer triplet with $m>1$.
Substituting the eigenvalue definitions into Eq.~\ref{eq:equal-args}, we see that the $n_m$th Bessel term in the $m$th region can be 
reused for the $n_e$th Bessel term in the exterior region if the frequency $\omega$ and index $n_e$ are set as follows:
\begin{equation}\label{eq:reuse-bessel-criterion}
    \omega=\sqrt{\frac{-g}{h}\pi\gamma \tan(\pi\gamma)},
    \qquad n_e = \lceil \gamma \rceil,
    \qquad  \gamma = n_m \frac{h}{h-d_m} \frac{a_m}{a_M}
\end{equation}
where we have introduced $\gamma$, a value that must be evaluated for each $(n_m,m)$ pair. Note that Eq.~\ref{eq:reuse-bessel-criterion} requires the tangent term to be negative, equivalently $0.5<\gamma-\lfloor \gamma \rfloor<1$, and any  $(n_m,m)$ pairs that do not meet this criteria cannot be reused. $\lceil \cdot\rceil$ and $\lfloor \cdot\rfloor$ denote the ceiling and floor functions respectively.

For the simple $M=2$ geometry described in section~\ref{sec:validation}, 70 of the 149 $\gamma$ values (47\%) are valid for reuse. 
However, only two of these (3\%) correspond to the 5-12 second wave periods typically of interest for ocean environments, so the computational savings are marginal. 
For the corresponding slant-discretized geometry with $M=16$, 1501 of the 2235 $\gamma$ values (51\%) are valid for reuse, including 64 (6\%) in the frequency range of interest. 
If this trend holds more broadly, then the aforementioned optimization may therefore be of particular utility to decrease the compute time for slanted geometries.
\begingroup
\begin{table}
\centering \begin{tabular}{|l|c|}
\hline
Function & Order of Time Complexity\\
\hline
Nonlinear solve for $\lambda_n^e$ & $N^e$\\
\hline
Coupling integrals ($\boldsymbol{\mathcal{Z}}^{i_m i_{m+1}}, \boldsymbol{\mathcal{Z}}^{i_M e}$) & $(\sum_{m=1}^{M-1} N^{i_m}N^{i_{m+1}}) + N^{i_M}N^e$\\
\hline
Bessel Functions in $\mathbf{A}$ *& $N^{i_1}+2\cdot(\sum_{m=2}^{M} N^{i_m}) + N^e$\\
\hline
Matrix solve ($\mathbf{A}\vec x^q = \vec b^q$) & $(N^{i_1}+2\cdot(\sum_{m=2}^{M} N^{i_m}) + N^e)^3$\\
\hline
Hydrodynamic Coefficients* & $N^{i_1}+2\cdot(\sum_{m=2}^{M} N^{i_m})$\\
\hline
\end{tabular}
\caption{Time complexities of major components of MEEM. All are represented as proportional to the number of times being called, except the matrix solve, which depends on the matrix size. \\ * The Bessel functions evaluations in $\vec c_p$ are the same as those in $\mathbf{A}$, so they are cached and their associated time ignored in measuring the time contributions of the $\vec c_p$ calculation.} \label{tab:time-complexities} \end{table}
\endgroup
\subsection{Comparison With Capytaine}
\label{sec:meem-cpt-timing}
\begin{figure}[htbp]
    \centering
    \includegraphics[width=0.9\linewidth]{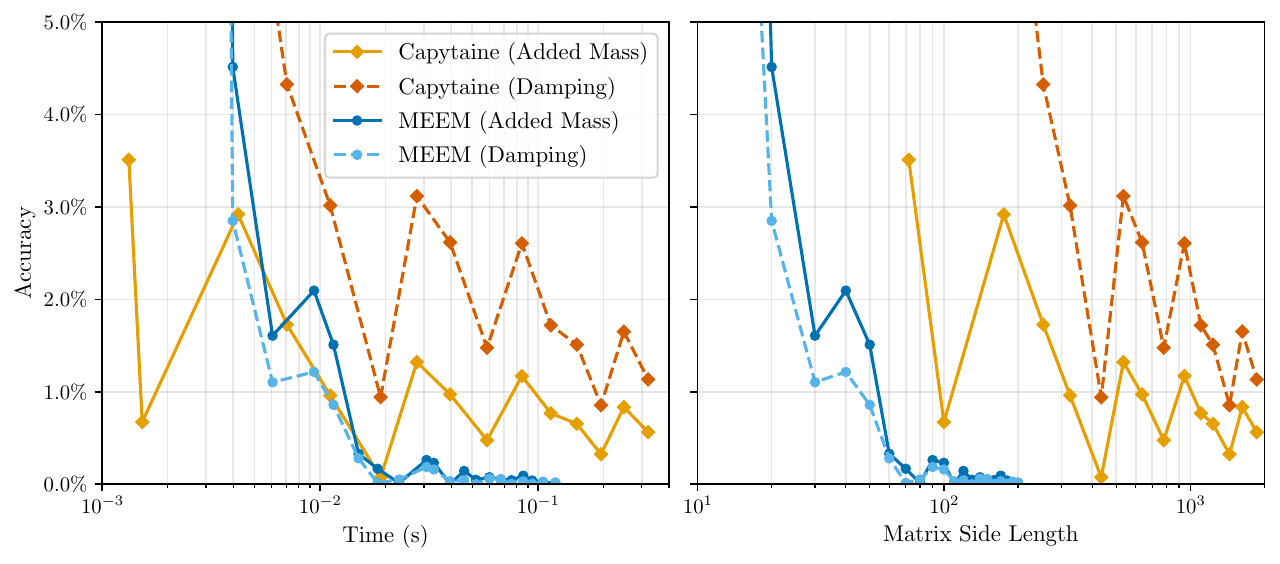}
    \caption{For the CorPower WEC geometry in Fig.~\ref{fig:hydro coeff validation} at $\omega = 1$, the accuracy of MEEM vs. Capytaine are compared over different $N^{i_m} = N^e$ and panel counts, through the associated computation time and matrix sizes. The ``true values'' the accuracies are relative to were determined by MEEM with $N^{i_m} = N^e = 300$. Figure context: \href{https://calkit.io/symbiotic-engineering/openflash/figures?path=pubs\%2FJFM\%2Ffigs\%2FMEEM-CPT-Time-Matrix-Comparison.pdf}{Calkit} and \href{https://cocalc.com/rmccabe/MEEM/meem}{CoCalc}.}
    \label{fig:time-matrix-comparison}
\end{figure}

For geometries that MEEM represents exactly, it tends to converge faster than Capytaine. An example of this is shown in Figure~\ref{fig:time-matrix-comparison}, where a CorPower-like WEC geometry is modelled at a single frequency and the trade-offs between accuracy and computation time (or matrix size) are compared. The added mass and radiation damping computed with MEEM can achieve $2\%$ accuracy with a runtime of $0.01$ seconds, while the values computed with Capytaine obtain $2\%$ accuracy after $0.1$ seconds. In terms of matrix size, Capytaine requires a matrix with a side length greater than $1000$ to achieve $2\%$ accuracy, while MEEM can achieve the same accuracy with a matrix side length of just $60$. This is a consequence of MEEMs fast convergence rate relative to BEM solvers' slow convergence rate with respect to the number of panels required to obtain an accurate solution. Such an advancement can enable more rigorous optimization studies for marine structures by avoiding the computational cost of BEM solvers.
%
\section{Conclusion}\label{sec:conclusion}
This work generalizes the matched eigenfunction expansion method (MEEM) for computing the hydrodynamic forces on an arbitrary number of heaving surface-piercing axisymmetric geometries under linear potential flow theory. The method consists of separating the fluid domain into cylindrical regions, defining the velocity potential in each region in terms of eigenfunctions and unknown eigencoefficients, and using boundary conditions and orthogonality to derive a system of linear algebraic equations, which can be organized into a matrix structure. This new extension overviews the undocumented accuracy, convergence, and polynomial runtime of the method. Furthermore, the ability of this method to be used to approximate the hydrodynamic forces on geometries with slants is discussed. This is done by subdividing slanted regions into cylindrical regions that can be modelled through MEEM. We have found that, while the velocity potential computed from MEEM may have inaccuracies locally, the hydrodynamic coefficients can be computed with less than 5\% error. Additionally, geometries with steeper slants require more subdivisions for MEEM to achieve accurate results. In the convergence study, a set of influential dimensionless parameters was identified, and the fitting models developed can predict the required settings of MEEM to achieve less than $2\%$ error in the hydrodynamic coefficients $95\%$ of the time. When compared against the boundary element method (BEM) solver Capytaine, MEEM is able to achieve 2\% convergence in both hydrodynamic coefficients an order of magnitude faster than Capytaine with a matrix size two orders of magnitude smaller, making it a computationally effective alternative to traditional BEM solvers. Such an advantage can have significant implications for the design optimization of marine structures, especially in situations when the hydrodynamic model is the bottleneck of the simulation model.

Further generalizations should be the focus of future work on this method.
Many geometric extensions have already been mathematically demonstrated for simple one-off cases, but they have yet to be integrated into a general multi-region matrix framework akin to the one presented here, nor have their numerical properties been examined.
Examples include degrees-of freedom beyond heave \citep{yeung_added_1981}, axisymmetric geometries for which the body radius varies non-monotonically with depth, as in the case of damping plates \citep{olaya_hydrodynamic_2015}, and axisymmetric bodies with discontinuous profiles creating so-called ``moon-pools'' \citep{seah_symmetric_2006}.
Equivalent analytical and semi-analytical potential flow solutions for other coordinate systems, such as Cartesian \citep{michele_theory_2016,renzi_hydrodynamics_2013}, elliptical \citep{nguyen_theoretical_2024,nguyen_theoretical_2024-1}, and spherical \citep{chatjigeorgiou_analytical_2018}, 
should be synthesized and unified with the present cylindrical approach.
To the authors' knowledge, no prior work has explored the use of MEEM for bodies of non-concentric cylinders, such as the OC4 semisubmersible floating offshore wind turbine, which could likely be accomplished via integration with the multiple scattering method. 
Each of these extensions would help ease the geometric restrictions that still plague MEEM compared to traditional BEM solvers.
Naturally, continued numerical comparison against BEM is necessary to evaluate the matrix properties, accuracy, and computation time as the framework evolves and expands to include more geometries.
It is the authors' hope that the OpenFLASH python package developed for this work can serve as a centralized repository for the software to implement such studies.

An additional avenue for further research is the practical application of the MEEM method, and especially of the insights gained from its mathematical structure and convergence behaviour exposed here, for hydrodynamic design.
\citet{mccabe_development_2026} describes the authors' initial attempt at such an endeavor, primarily leveraging MEEM's faster computational speed to widely explore the design space, and future work should implement the geometry-dependent number of terms heuristic from the convergence study in optimization.
Furthermore, MEEM's sparse linear solution structure could be leveraged in design by applying the adjoint method or other analytic or automatic differentiation techniques to enable scalable gradient-based optimization and sensitivity analysis of marine structures.
MEEM's smaller sparse matrices could help address the issues with large dense matrices encountered in the authors' BEM sensitivity work \citep{khanal_fully_2025}.

Finally, future work should investigate techniques similar to MEEM that exist across other fields.
Wave propagation boundary value problems are common in acoustics, electromagnetics, quantum mechanics, and seismology, 
and the notion of breaking a structured domain into analytically tractable subdomains has been tried in all of them to various degrees.
Understanding, comparing, and applying the insights from analogous problems in these diverse subjects could hasten the maturity of MEEM for hydrodynamics.
In particular, MEEM has striking similarity to the ``indirect Trefftz collocation method'' in finite element analysis.
This is a subset of the larger family of Trefftz methods, which refer to eigenfunctions as ``T-complete functions'' and also include least-square, Galerkin, variational, and direct versions \citep{kita_trefftz_1995}.
Exploring this connection could be especially relevant due to the large amount of attention that finite element analysis receives in the engineering and numerical computation communities.

All in all, MEEM holds computational promise as a result of its semi-analytical nature, sparse block structure, and rapid convergence properties.
\bibliographystyle{jfm}
\bibliography{jfm,becca-zotero-refs,zotero-meem-refs}





\begin{bmhead}[Code Availability.]
Code for all simulation, analysis, and visualization to fully reproduce this work is available open-source via the \texttt{OpenFLASH} project at \url{https://github.com/symbiotic-engineering/OpenFLASH} \citep{best_openflash_2026,best_openflash-code_2026}.
Questions and contributions via GitHub issues and pull requests are welcomed.
The Calkit and CoCalc links in the figure captions direct to Jupyter notebooks showing how figures were generated, allowing for replication and further exploration in a reproducible environment.
\end{bmhead}

\begin{bmhead}[Acknowledgements.]
We acknowledge Hope Best, Ruiyang Jiang, and John Fernandez who contributed to the software development of the \texttt{OpenFLASH} package.

Portions of the visualization and computational code used in this work were written with the help of generative AI tools.
This includes GitHub Copilot, Gemini 3 Pro, and ChatGPT using models available in 2025-26.
Authors reviewed and edited any AI-generated code to ensure accuracy.
With the exception of spelling and formatting suggestions, the authors performed all manuscript writing, analysis, and interpretation of results without the use of AI tools.

This paper is a heavily revised and expanded version of the authors' conference paper \citet{mccabe_open-source_2024}.
\end{bmhead}

\begin{bmhead}[Declaration of Interests.]
The authors report no conflict of interest.
\end{bmhead}

\begin{bmhead}[Funding Information.]
This work was supported in part by Sandia National Laboratories through the Marine Energy Seedlings Program for National Laboratories from the U.S. Department of Energy and Sandia's Laboratory Directed Research \& Development (LDRD) program through the Sandia University Partnerships Network. 
Sandia National Laboratories is a multimission laboratory managed and operated by National Technology \& Engineering Solutions of Sandia, LLC, a wholly owned subsidiary of Honeywell International Inc., for the U.S. Department of Energy's National Nuclear Security Administration under contract DE-NA0003525.

This material is based upon work supported by the National Science Foundation Graduate Research Fellowship under Grant No. DGE-2139899. Any opinion, findings, and conclusions or recommendations expressed in this material are those of the authors and do not necessarily reflect the views of the National Science Foundation.

Individual authors acknowledge the following sources of support:
Y. Bimali, Cornell Engineering Learning Initiatives through the Fund for Undergraduate Research on Solutions to Climate Change, the Bill Nye '77 Award in Undergraduate Research \& Robert A. Cowie '55 ME;
R. McCabe, National Science Foundation Graduate Research Fellowship Program, Cornell Provost's Diversity Fellowship, Cornell Mechanical and Aerospace Engineering McMullen Fellowship; 
C. Treacy, Sandia LDRD; 
K. Khanal, Sandia Seedlings program, Sandia LDRD; 
E. Lo, Cornell Engineering Learning Initiatives through the Dean Archer Undergraduate Research Program.

\end{bmhead}

\begin{bmhead}[Author ORCIDs.]{R. McCabe, https://orcid.org/0000-0001-5108-998X; C. Treacy, https://orcid.org/0009-0000-9381-2697; K. Khanal, https://orcid.org/0000-0002-0327-5945; E. Lo, https://orcid.org/0009-0005-0653-2715; M. Haji, https://orcid.org/0000-0002-2953-7253}
\end{bmhead}


\begin{bmhead}[Author contributions.]{Y. Bimali: Methodology, Software, Formal Analysis, Writing - original draft, review, and editing; R. McCabe: Conceptualization, Methodology, Software, Formal Analysis, Writing - original draft, review, and editing, Supervision, Project Administration; C. Treacy: Conceptualization, Methodology, Formal Analysis, Writing - original draft, review, and editing, Supervision, Project Administration; K. Khanal: Conceptualization, Methodology, Software, Formal Analysis, Writing - review and editing, Supervision, Project Administration; E. Lo: Methodology, Software, Formal Analysis, Writing - review and editing; M. Haji: Conceptualization, Writing - review and editing, Supervision, Project Administration, Funding Acquisition.}
\end{bmhead}
%
%
%
%
%
%
%
%
\begin{appen}\section{Governing Equations and Boundary Conditions}\label{appA}
The velocity potentials for the internal regions must satisfy
\begin{equation}\label{Laplace equation internal region}
    \nabla^2\phi^{i_m}=0,   
\end{equation}
\begin{equation}\label{No flux BC at wetted surface}
    \left. \frac{\partial\phi^{i_m}}{\partial z} \right|_{z=-d_m}=
    \begin{cases}
        0 & \text{ when region } m \text{ is fixed} \\
        1 & \text{ when region } m \text{ is heaving}
    \end{cases}
\end{equation}
and
\begin{equation}\label{No flux BC at sea floor internal}
    \left. \frac{\partial\phi^{i_m}}{\partial z} \right|_{z=-h}=0,   
\end{equation}
while the velocity potential for the external region must satisfy
\begin{equation}\label{Laplace equation external region}
    \nabla^2\phi^{e}=0,   
\end{equation}
\begin{equation}\label{Free surface BC}
    \left( -\frac{\omega^2}{g}\phi^{e} + \left. \frac{\partial\phi^{e}}{\partial z} \right) \right|_{z=0}= 0,
\end{equation}
and
\begin{equation}\label{No flux BC at sea floor external}
    \left. \frac{\partial\phi^{e}}{\partial z} \right|_{z=-h}=0.   
\end{equation}
The velocity potential in the interior regions can be written as the superposition of a homogeneous part $\phi^{i_m}_\mathrm{h}$, which corresponds to the first case of Eq.~\ref{No flux BC at wetted surface} when the body in region $m$ is fixed, and a particular part $\phi^{i_m}_\mathrm{p}$, which corresponds to the second case of Eq.~\ref{No flux BC at wetted surface} when the body in region $m$ is heaving with unit amplitude velocity.

The Laplace equations in Eq.~\ref{Laplace equation internal region} and~\ref{Laplace equation external region} can be solved using separation of variables in cylindrical coordinates. The velocity potentials can be written as a product of eigenfunctions $\phi (r, \theta, z)= R(r)\Theta(\theta)Z(z)$, where $\phi$ is $\phi^{i_m}$ and $\phi^{e}$ for the internal and external regions, respectively. Substituting this into the Laplace equation and separating each variable yields the following system of ordinary differential equations
\begin{equation}\label{eq:Z-ODE}
    \frac{\mathrm{d}^2 Z}{\mathrm{d}z^2}-\lambda^2Z=0
\end{equation}
\begin{equation}\label{eq:Theta-ODE}
    \frac{\mathrm{d}^2 \Theta}{\mathrm{d}\theta^2}+\nu^2\Theta=0
\end{equation}
\begin{equation}\label{eq:R-ODE}
    r^2\frac{\mathrm{d}^2 R}{\mathrm{d}r^2}+r\frac{\mathrm{d} R}{\mathrm{d}r} + (\lambda^2r^2-\nu^2)R=0
\end{equation}
where $R(r)$, $\Theta(\theta)$, and $Z(z)$ are eigenfunctions, and $\lambda$ and $\nu$ are eigenvalues. Since we are considering a vertically axisymmetric geometry and only heave motion, $\Theta=1$ and $\nu=0$. This leaves the vertical and radial eigenfunctions to be determined. As discussed in \citet{chatzigeorgiou2018analytical}, there are two cases of the eigenvalue $\lambda$ to consider: $\lambda \in \mathbb{R}$ and $\lambda \in \mathbb{I}$. In the first case, Eq.~\ref{eq:R-ODE} is the Bessel differential equation with Bessel and Hankel functions of the first and second kinds being solutions. In the second case, Eq.~\ref{eq:R-ODE} is the modified Bessel differential equation with modified Bessel functions of the first and second kinds being solutions.

The expression for $N_{n_e}$ is defined as:

\begin{equation}
    N_{n_e} = \frac{1}{2}\left(1+\frac{f_{n_e}}{2\lambda_{n_e}^eh}  \right)~
    \textrm{ where }
    f_{n_e} = 
    \begin{cases}
        \sinh(2 \lambda_0^e h), & n_e=0 \\ \sin(2\lambda_{n_e}^eh), & n_e\geq1
    \end{cases}.
\end{equation}
\section{Matching Equations}\label{appB}
Since both the value of the potential and fluid velocity must match at the boundary of neighboring regions, there are a total of $2M$ matching equations. However, these equations alone are not enough to solve for all eigencoefficients since the number of unknowns is greater than the number of equations ($N_\mathrm{T} >2M$). To generate an equal number of equations as unknowns, the orthogonality of the vertical eigenfunctions can be leveraged to isolate the unknown coefficients in the finite series. 




To derive a system of linear algebraic equations in terms of the eigencoefficients, we first multiply both sides of Eq.~\ref{potential matching} and~\ref{velocity matching} by a vertical eigenfunction, $Z^\mathrm{s}_{n_\mathrm{s}}(z)$ for potential matching and $Z^\mathrm{t}_{n_\mathrm{t}}(z)$ for velocity matching, and integrate over the fluid-fluid boundary to get
\begin{equation}\label{potential matching integral}
    \int_{-h}^{-d_\mathrm{s}}\phi^\mathrm{t}(a,z) Z^\mathrm{s}_{n_\mathrm{s}}(z) \mathrm{d}z=\int_{-h}^{-d_\mathrm{s}}\phi^\mathrm{s}(a,z) Z^\mathrm{s}_{n_\mathrm{s}}(z) \mathrm{d}z
\end{equation}
for potential matching and
\begin{equation}\label{velocity matching integral}
    \int_{-h}^{-d_\mathrm{s}}\frac{\partial \phi^\mathrm{t}}{\partial r}(a,z) Z^\mathrm{t}_{n_\mathrm{t}}(z) \mathrm{d}z=\int_{-h}^{-d_\mathrm{s}}\frac{\partial \phi^\mathrm{s}}{\partial r}(a,z) Z^\mathrm{t}_{n_\mathrm{t}}(z) \mathrm{d}z.
\end{equation}
for velocity matching.
The correct choice of vertical eigenfunction to multiply by in this step is subtle yet critical because it determines the side on which orthogonality will occur in later steps.
We want orthogonality to occur on the shorter side for potential matching because we have no information about the potential on the interval $z\in(-d_s,-d_t)$ which is included in the taller side's orthogonal domain.
On the other hand, we want orthogonality to occur on the taller side for velocity matching because we not only have information about the velocity on the interval $z\in(-d_s,-d_t)$ from boundary condition Eq.~\ref{no radial velocity}, but we have no other way to enforce this condition without incorporating it in the present equation.
Specifically, incorporating the condition can be accomplished because Eq.~\ref{no radial velocity} states that the integrand of the left-hand-side of Eq.~\ref{velocity matching integral} is zero in the interval $-d_\mathrm{s} \le z \le -d_\mathrm{t}$, allowing us to change the integration bounds on the left-hand-side of Eq.~\ref{velocity matching integral} from $-h \le z \le -d_\mathrm{s}$ to $-h \le z \le -d_\mathrm{t}$:
\begin{equation}\label{modified velocity matching integral}
    \int_{-h}^{-d_\mathrm{t}}\frac{\partial \phi^\mathrm{t}}{\partial r}(a,z)Z^\mathrm{t}_{n_\mathrm{t}}(z) \mathrm{d}z=\int_{-h}^{-d_\mathrm{s}}\frac{\partial \phi^\mathrm{s}}{\partial r}(a,z) Z^\mathrm{t}_{n_\mathrm{t}}(z) \mathrm{d}z.
\end{equation}
It is not desired to change the bounds of the right-hand-side because $\phi^s$ is not meaningfully defined on $z\in(-d_s,-d_t)$.
Thus, Eqs.~\ref{no radial velocity} and~\ref{velocity matching} are simultaneously enforced by Eq.~\ref{modified velocity matching integral}, while Eq.~\ref{potential matching} is enforced by Eq.~\ref{potential matching integral}. 

Next, the velocity potentials of each region can be rewritten in terms of their homogeneous and particular solutions as $\phi^\mathrm{t}=\phi^\mathrm{t}_\mathrm{h} + \phi^\mathrm{t}_\mathrm{p}$ and $\phi^\mathrm{s}=\phi^\mathrm{s}_\mathrm{h} + \phi^\mathrm{s}_\mathrm{p}$. Substituting these into Eq.~\ref{potential matching integral} and~\ref{modified velocity matching integral}, and moving all homogeneous and particular potentials to the left- and right-hand sides, respectively, yields
\begin{multline}\label{separated potential matching integral}
    \int_{-h}^{-d_\mathrm{s}} \phi^\mathrm{s}_\mathrm{h}(a,z) Z^\mathrm{s}_{n_\mathrm{s}}(z) \mathrm{d}z-\int_{-h}^{-d_\mathrm{s}} \phi^\mathrm{t}_\mathrm{h}(a,z) Z^\mathrm{s}_{n_\mathrm{s}}(z) \mathrm{d}z \\
    =\int_{-h}^{-d_\mathrm{s}}\phi^\mathrm{t}_\mathrm{p}(a,z) Z^\mathrm{s}_{n_\mathrm{s}}(z) \mathrm{d}z-\int_{-h}^{-d_\mathrm{s}}\phi^\mathrm{s}_\mathrm{p}(a,z) Z^\mathrm{s}_{n_\mathrm{s}}(z) \mathrm{d}z
\end{multline}
and
\begin{multline}\label{separated velocity matching integral}
    \int_{-h}^{-d_\mathrm{s}} \frac{\partial \phi^\mathrm{s}_\mathrm{h}}{\partial r}(a,z)  Z^\mathrm{t}_{n_\mathrm{t}}(z) \mathrm{d}z-\int_{-h}^{-d_\mathrm{t}} \frac{\partial \phi_\mathrm{h}^\mathrm{t}}{\partial r}(a,z) Z^\mathrm{t}_{n_\mathrm{t}}(z) \mathrm{d}z \\ 
    =\int_{-h}^{-d_\mathrm{t}}\frac{\partial \phi_\mathrm{p}^\mathrm{t}}{\partial r}(a,z) Z^\mathrm{t}_{n_\mathrm{t}}(z) \mathrm{d}z-\int_{-h}^{-d_\mathrm{s}}\frac{\partial \phi_\mathrm{p}^\mathrm{s}}{\partial r}(a,z) Z^\mathrm{t}_{n_\mathrm{t}}(z) \mathrm{d}z
\end{multline}
The right-hand-side of Eq.~\ref{separated potential matching integral} and~\ref{separated velocity matching integral} are known, while the left-hand-side contains unknowns. After substituting for the homogeneous potentials, using orthogonality of the vertical eigenfunctions and rearranging, Eq.~\ref{separated potential matching integral} and~\ref{separated velocity matching integral} become
\begin{multline}\label{separated potential matching integral simplified}
    (h-d_\mathrm{s}) \left( C^\mathrm{s}_{1n_\mathrm{s}} R_{1n_\mathrm{s}}^\mathrm{s} (a)+ C^\mathrm{s}_{2n_\mathrm{s}} R_{2n_\mathrm{s}}^\mathrm{s}(a) \right) \\ 
    -\sum _{n_\mathrm{t} = 0}^{N^\mathrm{t}} \left(C^\mathrm{t}_{1n_\mathrm{t}} R_{1n_\mathrm{t}}^\mathrm{t}(a)+ C^\mathrm{t}_{2n_\mathrm{t}} R_{2n_\mathrm{t}}^\mathrm{t}(a) \right)\int_{-h}^{-d_\mathrm{s}} Z^\mathrm{s}_{n_\mathrm{s}}(z) Z_{n_\mathrm{t}}^\mathrm{t}(z)\mathrm{d}z \\ 
    =\int_{-h}^{-d_\mathrm{s}}\left( \phi_\mathrm{p}^\mathrm{t}(a,z) - \phi_\mathrm{p}^\mathrm{s}(a,z) \right) Z^\mathrm{s}_{n_\mathrm{s}}(z) \mathrm{d}z
\end{multline}
and
\begin{multline}\label{separated velocity matching integral simplified}
    \sum _{n_\mathrm{s} = 0}^{N^\mathrm{s}} \left(C^\mathrm{s}_{1n_\mathrm{s}} \frac{\partial R_{1n_\mathrm{s}}^\mathrm{s}}{\partial r}(a)+ C^\mathrm{s}_{2n_\mathrm{s}} \frac{\partial R_{2n_\mathrm{s}}^\mathrm{s}}{\partial r}(a) \right)\int_{-h}^{-d_\mathrm{s}} Z^\mathrm{t}_{n_\mathrm{t}}(z) Z_{n_\mathrm{s}}^\mathrm{s}(z)  \mathrm{d}z \\ - (h-d_\mathrm{t}) \left( C^\mathrm{t}_{1n_\mathrm{t}} \frac{\partial R_{1n_\mathrm{t}}^\mathrm{t}}{\partial r} (a)+ C^\mathrm{t}_{2n_\mathrm{t}} \frac{\partial R_{2n_\mathrm{t}}^\mathrm{t}}{\partial r}(a) \right) \\ = \int_{-h}^{-d_\mathrm{t}}\frac{\partial \phi_\mathrm{p}^\mathrm{t}}{\partial r}(a,z) Z^\mathrm{t}_{n_\mathrm{t}}(z) \mathrm{d}z-\int_{-h}^{-d_\mathrm{s}}\frac{\partial \phi_\mathrm{p}^\mathrm{s}}{\partial r}(a,z) Z^\mathrm{t}_{n_\mathrm{t}}(z) \mathrm{d}z.
\end{multline}
Applying Eq.~\ref{separated potential matching integral simplified} for $n_\mathrm{s}=0,1,\dots, N^\mathrm{s}-1$ and Eq.~\ref{separated velocity matching integral simplified} for $n_\mathrm{t}=0,1,\dots, N^\mathrm{t}-1$ yields the matrix equations
\begin{multline}\label{potential matching vector equations}
    (h-d_\mathrm{s})\mathrm{diag}\left( \vec{R}_{1}^\mathrm{s}(a)\right) \vec{C}_{1}^\mathrm{s}  + (h-d_\mathrm{s})\mathrm{diag}\left(\vec{R}_{2}^\mathrm{s}(a)\right) \vec{C}_{2}^\mathrm{s} \\ - \boldsymbol{\mathcal{Z}}^{\mathrm{st}} \odot \boldsymbol{1}_{N^\mathrm{s}1}\vec{R}_{1}^\mathrm{t}(a)\vec{C}_1^\mathrm{t}  - \boldsymbol{\mathcal{Z}}^{\mathrm{st}} \odot \boldsymbol{1}_{N^\mathrm{s}1}\vec{R}_{2}^\mathrm{t}(a)\vec{C}_2^\mathrm{t} \\ =\int_{-h}^{-d_\mathrm{s}}\left( \phi_\mathrm{p}^\mathrm{t}(a,z) - \phi_\mathrm{p}^\mathrm{s}(a,z) \right) \vec{Z}^\mathrm{s}(z) \mathrm{d}z
\end{multline}
and
\begin{multline}\label{velocity matching vector equations}
    \boldsymbol{\mathcal{Z}}^\mathrm{ts} \odot \mathbf{1}_{N^\mathrm{t}1}\vec{R}_1^\mathrm{s}(a)\vec{C}_1^\mathrm{s}+\boldsymbol{\mathcal{Z}}^\mathrm{ts} \odot \mathbf{1}_{N^\mathrm{t}1}\vec{R}_2^\mathrm{s}(a)\vec{C}_2^\mathrm{s}\\ -(h-d_\mathrm{t})\mathrm{diag}\left( \frac{\partial}{\partial r} \vec{R}_1^\mathrm{t}(a)\right)\vec{C}_1^\mathrm{t} -(h-d_\mathrm{t})\mathrm{diag}\left( \frac{\partial}{\partial r} \vec{R}_2^\mathrm{t}(a)\right)\vec{C}_2^\mathrm{t} \\ = \int_{-h}^{-d_\mathrm{t}}\frac{\partial \phi_\mathrm{p}^\mathrm{t}}{\partial r}(a,z) \vec{Z}^\mathrm{t}{}(z) \mathrm{d}z-\int_{-h}^{-d_\mathrm{s}}\frac{\partial \phi_\mathrm{p}^\mathrm{s}}{\partial r}(a,z) \vec{Z}^\mathrm{t}{}(z) \mathrm{d}z
\end{multline}
where 
\begin{equation}
    \vec{R}_1^\ell(r)=[R_{10}^\ell(r), R_{11}^\ell(r),\dots, R_{1(N^\ell-1)}^\ell(r) ]
\end{equation}
and
\begin{equation}
     \vec{R}_2^\ell(r)=[R_{20}^\ell(r), R_{21}^\ell(r),\dots, R_{2(N^\ell-1)}^\ell(r) ] 
\end{equation}
 are row vectors of radial eigenfunctions;
\begin{equation}
    \vec{C}_1^\ell=[C_{10}^\ell, C_{11}^\ell,\dots, C_{1(N^\ell-1)}^\ell]^T
\end{equation}
and
\begin{equation}
    \vec{C}_2^\ell=[C_{20}^\ell, C_{21}^\ell,\dots, C_{2(N^\ell-1)}^\ell]^T
\end{equation}
 are column vectors of eigencoefficients; and
\begin{equation}
\begin{aligned}
    &\vec{Z}^\ell(z)=[Z_0^\ell(z),Z_1^\ell(z),\dots, Z_{(N^\ell-1)}^\ell(z)]^T 
\end{aligned}
\end{equation}
are column vectors of vertical eigenfunctions, where $\ell$ is $\mathrm{s}$ or $\mathrm{t}$. Additionally,  
\begin{equation}
    \begin{aligned}    &\boldsymbol{\mathcal{Z}}^\mathrm{st}=\boldsymbol{\mathcal{Z}}^\mathrm{ts}{}^T=\int_{-h}^{-d_\mathrm{s}} \vec{Z}^\mathrm{s}{}(z) \otimes \vec{Z}^\mathrm{t}(z) \mathrm{d}z
\end{aligned}
\end{equation}
is an $N^\mathrm{s} \times N^\mathrm{t}$ matrix of coupling integrals; $\mathbf{1}_{N^\mathrm{s}1}$ and $\mathbf{1}_{N^\mathrm{t}1}$ are $N^\mathrm{s} \times 1$ and $N^\mathrm{t} \times 1$ matrices of ones, respectively; $\odot$ is the Hadamard (element-wise) product; and $\otimes$ is outer product.

  Eq.~\ref{potential matching vector equations} and~\ref{velocity matching vector equations} contain $N^\mathrm{s}$ and $N^\mathrm{t}$ equations, respectively. However, they have $2N^\mathrm{s} + 2N^\mathrm{t}$ unknown eigencoefficients. Recall that these equations correspond to enforcing conditions at a single boundary between fluid regions, as shown in Fig.~\ref{fig:Matching Diagram}. Section~\ref{Block Matrix Structure} shows how to organize Eq.~\ref{potential matching vector equations} and~\ref{velocity matching vector equations} for the $M$ boundaries so that all eigencoefficients can be solved for simultaneously.
  Because the innermost and outermost regions have just one eigenfunction instead of two and therefore half as many eigencoefficients, the full system has exactly as many equations as unknowns.

Note that all coupling integrals have closed form solutions. The element in the $n_\mathrm{s}$th row and $n_\mathrm{t}$th column of the coupling integral matrix associated with matching between region $i_m$ and $i_{m+1}$ is

\begin{equation}\label{coupling integral i_m and i_m+1}
    [\boldsymbol{\mathcal{Z}}^{\text{s} \text{t}} ]_{n_\mathrm{s}n_\mathrm{t}}=-\frac{\sin ((d_\text{s}-h) (\lambda_{n_\mathrm{s}}^{\text{s}}-\lambda_{n_\mathrm{t}}^{\text{t}}))}{\lambda_{n_\mathrm{s}}^{\text{s}}-\lambda_{n_\mathrm{t}}^{\text{t}}}-\frac{\sin ((d_\text{s}-h) (\lambda_{n_\mathrm{s}}^{\text{s}}+\lambda_{n_\mathrm{t}}^{\text{t}}))}{\lambda_{n_\mathrm{s}}^{\text{s}}+\lambda_{n_\mathrm{t}}^{\text{t}}},
\end{equation}

\noindent where the superscripts and subscripts associated with the shorter and taller fluid are assigned to $i_m$ or $i_{m+1}$ depending on if $d_m>d_{m+1}$ or $d_{m+1}>d_m$. Eq.~\ref{coupling integral i_m and i_m+1} holds for $1 \le m \le M-1$. For matching at the interface between the outermost inner region $i_M$ and the external region $e$, a different expression is used. The element in the $n_M$th row and $n_e$th column of the coupling integral matrix associated with matching between region $i_M$ and $e$ is
\begin{equation}\label{coupling integral i_M and e}
    [\boldsymbol{\mathcal{Z}}^{i_M e} ]_{n_Mn_e}=
    - \sqrt{\frac{1}{2N_{n_e}}} 
    \left( 
        \frac{\sin ((d_M-h) (\lambda_{n_e}^{e}-\lambda_{n_M}^{i_M}))}
        {(\lambda_{n_e}^{e}-\lambda_{n_M}^{i_M})} 
        + \frac{\sin ((d_M-h) (\lambda_{n_e}^{e}+\lambda_{n_M}^{i_M}))}
        {(\lambda_{n_e}^{e}+\lambda_{n_M}^{i_M})}
    \right).
\end{equation}
Note that Eq.~\ref{coupling integral i_m and i_m+1} and~\ref{coupling integral i_M and e} can simplify considerably when one or both indices are zero.

Rewriting Eq.~\ref{potential matching vector equations} and~\ref{velocity matching vector equations} for each of the $M$ boundaries, where the symbols representing the smaller $\mathrm{s}$ and taller $\mathrm{t}$ regions are replaced by the region names of the problem (i.e. $i_1, i_2,\dots,  i_M,e$) and $a$ is replaced by $a_m$, yields a set of $N_\mathrm{T}$ equations that are linear with respect to the unknown eigencoefficients.
The structure of these equations is in the form of $\mathbf{A} \vec{x}=\vec{b}$ where the matrix $\mathbf{A} \in \mathbb{C}^{N_\mathrm{T}\times N_\mathrm{T}}$ contains the left-hand-side of Eq.~\ref{potential matching vector equations} and~\ref{velocity matching vector equations} (excluding the eigencoefficients), the vector $\vec{x}=[\vec{C}_{1}^{i_1}, \vec{C}_{1}^{i_2}, \vec{C}_{2}^{i_2},\dots, \vec{C}_{1}^{i_M}, \vec{C}_{2}^{i_M}, \vec{C}_{1}^{e}]^T \in \mathbb{C}^{N_\mathrm{T}}$ contains all eigencoefficients, and the vector $\vec{b} \in \mathbb{R}^{N_\mathrm{T}}$ contains the right-hand-side of Eq.~\ref{potential matching vector equations} and~\ref{velocity matching vector equations}.
As shown in Table~\ref{tab:MEEM-A-matrix}, $\mathbf{A}$ is a block bi-diagonal matrix composed of sub-matrices $\mathbf{A}_1,\mathbf{A}_2,\dots, \mathbf{A}_M$ and zero matrices. Each sub-matrix $\mathbf{A}_m$ contains the left-hand-sides of Eq.~\ref{potential matching vector equations} and~\ref{velocity matching vector equations} (excluding the eigencoefficients) when applying them to the $m$th boundary.
The corresponding right-hand-sides of Eq.~\ref{potential matching vector equations} and~\ref{velocity matching vector equations} are contained in $\vec{b}_m$, which form the vector $\vec{b}=[\vec{b}_1,\vec{b}_2,\dots, \vec{b}_M]^T$. 
\section{Forms of Matrix A}\label{sec:Forms of Matrix A}
The matrix $\mathbf{A}_m$ can be written as 
$\mathbf{A}_m = \mathbf{A}_{m,r} \odot \mathbf{A}_{m,z}$, 
where $\mathbf{A}_{m,r}$ and $\mathbf{A}_{m,z}$ are shown in 
Tables~\ref{tab:MEEM-A_mr-matrix-case-1}-\ref{tab:MEEM-A_mz-matrix-case-2} 
and $\mathbf{I}^{i_m}$ is a $N^{i_m} \times N^{i_m}$ identity matrix.
\begin{landscape}
\begin{table}
    \centering
    \begin{tabular}{|>{\centering\arraybackslash}p{0.18\linewidth}|c||c|c|c|c|c|}
    \hline
    & & 
    $\vec{C}_{1}^{i_m}$
    & 
    $\vec{C}_{2}^{i_m}$
    & 
    $\vec{C}_{1}^{i_{m+1}}$ 
    & 
    $\vec{C}_{2}^{i_{m+1}}$ 
    \\\hline 
    &
    size
    &  
    $N^{i_m}$
    &  
    $N^{i_m}$
    &  
    $N^{i_{m+1}}$
    & 
    $N^{i_{m+1}}$
    \\ \hline \hline 
      
      \shortstack{
        $\phi^{i_m}=\phi^{i_{m+1}}$ 
        \\ at $r=a_m$} 
        & 
        $N^{i_m}$ 
        & 
        $\mathbf{1}_{N^{i_m}1} \vec{R}_1^{i_m}$ 
        & 
        $\mathbf{1}_{N^{i_m}1}  \vec{R}_2^{i_m}$ 
        & 
        $ \mathbf{1}_{N^{i_m}1} \vec{R}_1^{i_{m+1}}$ 
        & 
        $ \mathbf{1}_{N^{i_m}1} \vec{R}_2^{i_{m+1}}$ 
        \\ \hline
      
      \shortstack{
        $\frac{\partial}{\partial r}\phi^{i_m}=\frac{\partial}{\partial r}\phi^{i_{m+1}}$ 
        \\ at $r=a_m$} 
        & 
        $N^{i_{m+1}}$
        & 
        $ \mathbf{1}_{N^{i_{m+1}}1} \vec{R}_1^{i_{m}}$ 
        & 
        $ \mathbf{1}_{N^{i_{m+1}}1} \vec{R}_2^{i_{m}}$ 
        & 
        $\mathbf{1}_{N^{i_{m+1}}1}  \frac{\partial}{\partial r} \vec{R}_1^{i_{m+1}}$ 
        & 
        $\mathbf{1}_{N^{i_{m+1}}1}  \frac{\partial}{\partial r} \vec{R}_2^{i_{m+1}}$ 
        \\ \hline
    \end{tabular}
    \caption{MEEM $\mathbf{A}_{m,r}$ sub-matrix when $d_m>d_{m+1}$ ($i_m = \mathrm{s}$ and $i_{m+1} = \mathrm{t}$). 
    Note all radial eigenfunctions and their derivatives are evaluated at $r=a_m$.}
    \label{tab:MEEM-A_mr-matrix-case-1}
\end{table}
\end{landscape}
\begin{landscape}
\begin{table}
    \centering
    \begin{tabular}{|>{\centering\arraybackslash}p{0.18\linewidth}|c||c|c|c|c|c|}
    \hline
     & & 
    $\vec{C}_{1}^{i_m}$
    & 
    $\vec{C}_{2}^{i_m}$
    & 
    $\vec{C}_{1}^{i_{m+1}}$ 
    & 
    $\vec{C}_{2}^{i_{m+1}}$ 
    \\\hline 
    &size&  
    $N^{i_m}$
    &  
    $N^{i_m}$
    &  
    $N^{i_{m+1}}$
    & 
    $N^{i_{m+1}}$
    \\ \hline \hline 
      
    \shortstack{
                $\phi^{i_m}=\phi^{i_{m+1}}$ 
                \\ 
                at $r=a_m$
                } 
    & 
    $N^{i_{m+1}}$ 
    & 
    $ \mathbf{1}_{N^{i_{m+1}}1} \vec{R}_1^{i_{m}}$ 
    & 
    $\mathbf{1}_{N^{i_{m+1}}1} \vec{R}_2^{i_{m}}$ 
    & 
    $\mathbf{1}_{N^{i_{m+1}}1} \vec{R}_1^{i_{m+1}}$ 
    & 
    $ \mathbf{1}_{N^{i_{m+1}}1}  \vec{R}_2^{i_{m+1}}$ 
    \\ \hline
      
      \shortstack{
                    $\frac{\partial}{\partial r}\phi^{i_m}=\frac{\partial}{\partial r}\phi^{i_{m+1}}$ 
                    \\ 
                    at $r=a_m$
                } 
      & 
      $N^{i_{m}}$
      & 
      $\mathbf{1}_{N^{i_{m}}1} \frac{\partial}{\partial r} \vec{R}_1^{i_{m}}$ 
      & 
      $\mathbf{1}_{N^{i_{m}}1} \frac{\partial}{\partial r} \vec{R}_2^{i_{m}}$ 
      & 
      $ \mathbf{1}_{N^{i_{m}}1} \vec{R}_1^{i_{m+1}}$ 
      & 
      $ \mathbf{1}_{N^{i_{m}}1} \vec{R}_2^{i_{m+1}}$ 
      \\ \hline
    \end{tabular}
    \caption{MEEM $\mathbf{A}_{m,r}$ sub-matrix when $d_m<d_{m+1}$ ($i_m = \mathrm{t}$ and $i_{m+1} = \mathrm{s}$). 
    Note all radial eigenfunctions and their derivatives are evaluated at $r=a_m$.
    }
    \label{tab:MEEM-A_mr-matrix-case-2}
\end{table}
\end{landscape}
\begin{landscape}
\begin{table}
    \centering
    \begin{tabular}{|>{\centering\arraybackslash}p{0.18\linewidth}|c||c|c|c|c|c|}
    \hline
        &
        & 
        $\vec{C}_{1}^{i_m}$
        & 
        $\vec{C}_{2}^{i_m}$
        & 
        $\vec{C}_{1}^{i_{m+1}}$ 
        & 
        $\vec{C}_{2}^{i_{m+1}}$ 
    \\\hline 
        &
        size
        &  
        $N^{i_m}$
        &  
        $N^{i_m}$
        &  
        $N^{i_{m+1}}$
        & 
        $N^{i_{m+1}}$
    \\ \hline \hline 
      
      \shortstack{
        $\phi^{i_m}=\phi^{i_{m+1}}$ 
        \\ 
        at $r=a_m$} 
        & 
        $N^{i_m}$ 
        & 
        $(h-d_m) \mathbf{I}^{i_m}$ 
        & 
        $(h-d_m) \mathbf{I}^{i_m}$ 
        & 
        $-\boldsymbol{\mathcal{Z}}^{i_mi_{m+1}}$ 
        & 
        $-\boldsymbol{\mathcal{Z}}^{i_mi_{m+1}}$ 
        \\ \hline
      
      \shortstack{
            $\frac{\partial}{\partial r}\phi^{i_m}=\frac{\partial}{\partial r}\phi^{i_{m+1}}$ 
            \\ 
            at $r=a_m$
        } 
        & 
        $N^{i_{m+1}}$
        & 
        $\boldsymbol{\mathcal{Z}}^{i_{m+1}i_m}$ 
        & 
        $\boldsymbol{\mathcal{Z}}^{i_{m+1}i_m}$ 
        & 
        $-(h-d_{m+1}) \mathbf{I}^{i_{m+1}}$ 
        & 
        $-(h-d_{m+1}) \mathbf{I}^{i_{m+1}}$ 
        \\ \hline
    \end{tabular}
    \caption{MEEM $\mathbf{A}_{m,z}$ sub-matrix when $d_m>d_{m+1}$ ($i_m = \mathrm{s}$ and $i_{m+1} = \mathrm{t}$). 
    }
    \label{tab:MEEM-A_mz-matrix-case-1}
\end{table}
\end{landscape}
\begin{landscape}
\begin{table}
    \centering
    \begin{tabular}{|>{\centering\arraybackslash}p{0.18\linewidth}|c||c|c|c|c|c|}
    \hline
        & & 
        $\vec{C}_{1}^{i_m}$
        & 
        $\vec{C}_{2}^{i_m}$
        & 
        $\vec{C}_{1}^{i_{m+1}}$ 
        & 
        $\vec{C}_{2}^{i_{m+1}}$ 
    \\\hline 
        &
        size
        &  
        $N^{i_m}$
        &  
        $N^{i_m}$
        &  
        $N^{i_{m+1}}$
        & 
        $N^{i_{m+1}}$
    \\ \hline \hline 
      
      \shortstack{
        $\phi^{i_m}=\phi^{i_{m+1}}$ 
        \\ 
        at $r=a_m$} 
        & 
        $N^{i_{m+1}}$ 
        & 
        $-\boldsymbol{\mathcal{Z}}^{i_{m+1}i_m} $ 
        & 
        $-\boldsymbol{\mathcal{Z}}^{i_{m+1}i_m}$ 
        & 
        $(h-d_{m+1}) \mathbf{I}^{i_{m+1}}$ 
        & 
        $(h-d_{m+1}) \mathbf{I}^{i_{m+1}}$ 
    \\ \hline
      
      \shortstack{
        $\frac{\partial}{\partial r}\phi^{i_m}=\frac{\partial}{\partial r}\phi^{i_{m+1}}$ 
        \\ 
        at $r=a_m$} 
        & 
        $N^{i_{m}}$
        & 
        $-(h-d_{m}) \mathbf{I}^{i_{m}}$ 
        & 
        $-(h-d_{m}) \mathbf{I}^{i_{m}}$ 
        & 
        $\boldsymbol{\mathcal{Z}}^{i_mi_{m+1}} $ 
        & 
        $\boldsymbol{\mathcal{Z}}^{i_mi_{m+1}} $ 
        \\ \hline
    \end{tabular}
    \caption{MEEM $\mathbf{A}_{m,z}$ sub-matrix when $d_m<d_{m+1}$ ($i_m = \mathrm{t}$ and $i_{m+1} = \mathrm{s}$).}
    \label{tab:MEEM-A_mz-matrix-case-2}
\end{table}
\end{landscape}
\section{Added Mass and Damping}\label{appC}
We will define the velocity of the $q$th body as 
$v_q(t) = \mathrm{Re} \{ \hat{v}_q e^{- \text{i} \omega t}\}$ 
so that $\hat{f}_{pq} = (\text{i} \omega A_{pq}(\omega) - B_{pq}(\omega)) \hat{v}_q$. 
Recall from Eq.~\ref{No flux BC at wetted surface} that, if a region is heaving, 
the particular potentials in Table \ref{tab:MEEM-eigenfunctions} were defined so that it does so with unit amplitude velocity. 
Thus, if regions consisting of the $q$th body are heaving, $\hat{v}_q=1$. 
Eq.~\ref{eq:freq domain scalar rad force in terms of regions} can be simplified and rearranged as
\begin{equation}\label{eq:freq domain scalar rad force in terms of added mass and damping}
\begin{aligned}
    A_{pq}(\omega) + \frac{\text{i}B_{pq}(\omega)}{\omega} 
    &= 2 \pi \rho \sum_{m \in \mathcal{M}_p} \int_{a_m}^{a_{m+1}} \ _{}^{q}\phi^{i_m}(r,-d_m) \ r  \ dr \\
    &= 2 \pi \rho \sum_{m \in \mathcal{M}_p} \Bigg[
    \int_{a_{m}}^{a_{{m+1}}} \ _{}^{q}\phi^{i_m}_p(r,-d_m)\, r\, dr \\
    & + \sum_{n_m=0}^{N_{i_m}-1} 
    \ _{}^{q}C_{1n_m}^{i_m} \ Z_{n_m}^{i_m}(-d_m)
    \int_{a_{m}}^{a_{{m+1}}} \ R_{1n_m}^{i_m}(r)\, r\, dr \\
    & + \sum_{n_m=0}^{N_{i_m}-1} 
    \ _{}^{q}C_{2n_m}^{i_m} \ Z_{n_m}^{i_m}(-d_m)
    \int_{a_{m}}^{a_{{m+1}}} \ R_{2n_m}^{i_m}(r)\, r\, dr
    \Bigg]
\end{aligned}
\end{equation}
where all symbols written with a left superscript $q$ denote 
symbols specific to solving the radiation problem when only the $q$th body is moving.

Additionally, the $n_m$th element of the vectors 
$\vec{\mathcal{R}}_{1}^{i_m}$ and $\vec{\mathcal{R}}_{2}^{i_m}$ 
in Eq.~\ref{eq:c_q_i1_vec} and \ref{eq:c_q_im_vec} are 
\begin{equation}\label{radial integral 1}
\begin{aligned}\relax
    [\vec{\mathcal{R}}_{1}^{i_m}]_{n_m} 
    & = \int_{a_{m}}^{a_{{m+1}}} R_{1n_m}^{i_m}(r)\, r\, dr \\
    & = \left\{\begin{matrix}
    \frac{1}{4} (a_{{m+1}}^2-a_{m}^2) & \text{ for } n_m=0 \\[2ex]
 \frac{a_{{m+1}} \mathrm{I}_1(a_{{m+1}} \lambda_{n_m}^{i_m})-a_{m} \mathrm{I}_1(a_{m} \lambda_{n_m}^{i_m})}{\lambda_{n_m}^{i_m}\mathrm{I}_0(a_{{m}}\lambda_{n_m}^{i_m})} 
 & \text{ for } n_m \ge 1
\end{matrix}\right.
\end{aligned}
\end{equation}

\noindent and

\begin{equation}\label{radial integral 2}
\begin{aligned}\relax
    [\vec{\mathcal{R}}_{2}^{i_m}]_{n_m} 
    & = \int_{a_{m}}^{a_{{m+1}}} R_{2n_m}^{i_m}(r)\, r\, dr \\
    & = \left\{\begin{matrix}
    \frac{1}{8} \left(2 a_{{m+1}}^2 \ln \left(\frac{a_{{m+1}}}{a_{m}}\right)-a_{{m+1}}^2+a_{m}^2\right) 
    & 
    \text{ for } n_m=0 \\[2ex]
 \frac{
        a_m \mathrm{K}_1(a_m \lambda^{i_m}_{n_m})-a_{{m+1}} \mathrm{K}_1(a_{{m+1}} \lambda^{i_m}_{n_m})
        }{
        \lambda^{i_m}_{n_m} \mathrm{K}_0(a_m \lambda^{i_m}_{n_m})
        } 
 &
  \text{ for } n_m \ge 1
\end{matrix}.\right.
\end{aligned}
\end{equation}

\end{appen}\clearpage

\end{document}